\documentclass{emulateapj}
\usepackage{natbib}
\usepackage{epsfig,epsf}

\def\chisq{$\chi^2$}
\def\ergs{${\rm erg\,s^{-1}\,cm^{-2}}$}
\def\ulum{${\rm erg\,s^{-1}}$}
\def\mbh{$M_{\rm bh}$}

\def\msun{\ensuremath{M_{\odot}}}

\def\nh{$N_{\rm H}$}
\def\gnh{$N_{\rm H}^{\rm Gal}$}
\def\unh{${\rm cm^{-2}}$}
\def\kmps{${\rm km\,s^{-1}}$ }
\def\ucr{${\rm counts\,s^{-1}}$}
\def\redd{$L_{\rm bol}/L_{\rm Edd}$}
\def\reddx{$L_{\rm bol}^{\rm x}/L_{\rm Edd}$}

\def\hb{H$\beta$}
\def\ha{H$\alpha$}

\def\loiii{$L_{\rm [OIII]}$}
\def\lx{$L_{\rm x}$}
\def\lomon{$L_{\rm 2500\AA}$}
\def\pnull{$P_{\rm null}$}

\def\aox{$\alpha_{\rm ox}$}

\def\lb{$L_{\rm bol}$}

\newcommand{\nii}{[N\,{\footnotesize II}]}

\slugcomment{} 
\shorttitle{Low-mass BH accreting at low rates}
\shortauthors{Yuan W. et al.}

\begin{document}

\title{
Chandra and MMT observations of low-mass black hole
active galactic nuclei accreting at low rates 
in dwarf galaxies
}

\author{
W. Yuan\altaffilmark{1}, 
H. Zhou\altaffilmark{2,3},
L. Dou\altaffilmark{2},
X.-B. Dong\altaffilmark{2,4,5},
X. Fan\altaffilmark{6},
T.-G. Wang\altaffilmark{2}
}

\altaffiltext{1}{National Astronomical Observatories, 
Space Science Division, 
Chinese Academy of Sciences, Beijing, 100012, China; wmy@nao.cas.cn}

\altaffiltext{2}{Key Laboratory for Research in Galaxies and
Cosmology, Department of Astronomy, The University of Sciences and
Technology of China, Chinese Academy of Sciences, Hefei, Anhui
230026, China}

\altaffiltext{3}{Polar Research Institute of China, Jinqiao Road 451,
Shanghai 200136, China}

\altaffiltext{4}{Yunnan  Observatory, 
Chinese Academy of Sciences, Kunming, Yunnan 650011, China}

\altaffiltext{5}{Key Laboratory for the Structure and Evolution of
Celestial Objects, Chinese Academy of Sciences, Kunming, Yunnan 650011, China}

\altaffiltext{6}{Steward Observatory, The University of Arizona,
Tucson, AZ 85721, USA} 


\begin{abstract}

We report on Chandra X-ray observations of  
four candidate low-mass black hole (\mbh$\la10^6$\,\msun)
active galactic nuclei (AGNs) that have 
the estimated Eddington ratios among the lowest ($\sim10^{-2}$)  
found for this class.
The aims are to validate the nature of their AGNs  
and to confirm the low Eddington ratios that are derived
 from the broad \ha\ line, 
 and to explore this poorly studied regime  in the AGN parameter space.
Among them, two objects with the lowest significance of the broad lines
are also observed with Multi-Mirror Telescope, and the high-quality optical spectra taken
confirm them as Seyfert\,1 AGNs and as having small black hole masses.
X-ray emission is detected from the nuclei of two of the galaxies,
which is variable on timescales of $\sim 10^3$\,s,
whereas no significant  (or only marginal at best) detection 
is found for the remaining two. 
The X-ray luminosities are on the order of $10^{41}$\,\ulum\
or even lower, on the order of $10^{40}$\,\ulum\ for  non-detections,
which are among the lowest regimes ever probed for Seyfert galaxies.
The  low X-ray luminosities,
compared to their black hole masses derived from \ha, 
confirm their low accretion rates assuming typical bolometric corrections.
Our results hint at the existence of a possibly large population of under-luminous 
low-mass black holes
 in the local universe.
An off-nucleus ultra-luminous X-ray source 
in one of the dwarf galaxies is detected serendipitously, 
with a  luminosity  (6--9)$\times10^{39}$\,\ulum\ in 2--10\,keV.
\end{abstract}


\section{Introduction}

Black holes (BHs)  with masses in the range of \mbh$\sim 10^{3-6}\,M_{\odot}$,
termed 
intermediate-mass BHs (IMBHs),
are an astrophysically and cosmologically important 
class linking stellar mass BHs and massive/supermassive BHs
at the center of galaxies. 
They can be found by virtue of the accretion-powered radiation
shining as active galactic nuclei (AGNs) in nearby small/dwarf galaxies, 
albeit much less luminous than their more massive counterparts.
However, only a few hundred IMBHs are known so far, and very few 
have been studied in detail.\footnote{
These include, e.g.,\ NGC\,4395---the prototype and the best-studied 
of this type  \citep[e.g.][]{fhs93,lira99,fh03},
POX\,52 \citep{bar04} and SDSS\,J160531.84+174826.1 \citep{dong07}
}
Their apparent rarity, as indicated by the turnover of the
(observed) mass function of BHs in AGNs toward the lower end \citep{gh07a},
 raises a question as to whether they are 
truly scarce in the universe or there exists a large population
yet to be discovered. 
In another word, whether the majority of 
dwarf galaxies---the most abundant galactic population and
building blocks of large galaxies---harbor a BH at their centers. 
The answer, whatever it is, would have significant
implications for testing and constraining models of the birth and growth of 
galaxies and BHs \citep[e.g.][]{vol08}.

The first samples of AGNs with BH masses at the lower end of 
the mass function  ($\sim10^{5-6}$\,\msun) 
\footnote{Following \citet{gh07b} and \citet{dong12}, 
we refer to such BHs as low-mass BHs and to AGNs hosting them as low-mass AGNs
throughout the paper.
}
were found by Greene \& Ho (2004, 2007b) from the Sloan Digital Sky Survey (SDSS).
Most of those AGNs have (relatively) high Eddington ratios (\redd$>0.1$),
based on BH masses estimated from the broad \ha\ line.
The apparent dearth of (relatively) low-accretion low-mass AGNs (\redd$<0.1$) 
naturally explains the observed drop in 
the BH mass function below $\sim10^{6}$\,\msun. 
It is not yet settled by observations that, whether there is 
a true decline or even a cutoff in the BH mass function
as discussed theoretically 
(e.g.\ a few times $10^5$\,\msun\ as in Haehnelt et al.\ 1998),
or if this is simply a selection bias.
The latter comes into play since  low-mass AGNs accreting at low rates
are difficult to detect for the extreme faintness of
their continuum and line luminosities
(e.g. an AGN with \mbh=$10^6$\,\msun\ has a bolometric luminosity of 
$1.26\times 10^{42}$\,\ulum\ for \redd=0.01),
and the dilution by host galaxy starlight further aggravates the situation. 
To bear this out,  low-accretion-rate, low-mass AGNs are
to be sought out observationally as a population.
 
Few low-mass AGNs have been found with reliably measured low \redd\ thus far, 
however, because the multi-waveband spectral energy distribution (SED) 
is difficult to observe.
The best case known is NGC\,4395, which happens 
to be a low-Eddington ratio object with 
\redd$\approx 10^{-3}-10^{-2}$ depending on the measured \mbh\ value 
\citep[e.g.][]{fh03,pet05,mor05,iwa10},
which is possibly the lowest among low-mass AGNs observed\footnote{
This is in contrast to Seyfert galaxies 
with supermassive BHs ($\sim10^{7-9}$\,\msun),
in which \redd\ as low as $\la 10^{-3}$ were found \citep{ho09,sing11}.
In fact, in homogeneously selected AGN samples the {\em observed} 
lower bound of the \redd\ distribution increases with decreasing \mbh,
which may be explained by selection biases (see Dong X.\ et al.\ 2012).
}.
It is the least luminous Seyfert\,1 nucleus \citep{fhs93},
with \lb\ of the order of $10^{40}$\,\ulum\ based on the multi-waveband SED,
which happens to be comparable to the empirical estimate from 
the monochromatic luminosity \citep{iwa10}.
It was detected by virtue of its proximity and 
the absence of a galactic bulge, hence the reduced diluting starlight.
There are more candidates in the less stringent sense where 
the bolometric luminosities are estimated from only one waveband.
A small number of low-mass AGNs with \redd$\approx10^{-2}-10^{-1}$ estimated from \ha\
have been found in the \citet{gh07b} sample.
A few more possible candidates may have also been found 
in the search of weak AGN activity
in nearby (small) galaxies in other wavebands (X-ray, infrared, radio)
in some of the recent studies \citep{dh09,sat08}, 
but their Eddington ratios are subject to large uncertainties due to a 
lack of reasonable estimation of \mbh.

In a recent search of low-mass AGNs designed to be more complete and homogeneous,
we \citep{dong12} presented a sample of 309 objects (hereafter D12 sample)
selected from the forth SDSS data release (DR4), 
finding more low-accreting systems down to 
\redd$\sim 0.01$, and $<0.1$ in 30\% of the objects.
If the \redd\ estimates, which are solely
based on the broad \ha\ line luminosities 
(the \mbh\ values were estimated from the broad \ha\ line widths and
luminosities using the same formalism as in Greene \& Ho 2007b), 
are largely correct, 
this finding has at least two implications.
First, the presence of a postulated population of 
low-mass BHs accreting at low rates 
can thus be anchored to observations.
This would naturally hint at the possible
 existence of a continuous distribution in the
accretion rate for low-mass BHs and, hence, 
a population of low-mass BHs with weak activity.
Second, concerning theory and observation of BH accretion physics,
this regime in the parameter space is largely unexplored.
They may provide a clue that may reconcile some of the observed
differences between massive BSs and stellar-mass BHs
in X-ray binaries.

However, the estimation of the Eddington ratios based on \ha\
(especially for low-\redd\ objects) may be susceptible to some effects,
and thus independent scrutiny is essential.
First, the measurement of the broad H$\alpha$ component in
low-\redd\  systems (hence low \ha\ luminosity) 
has large uncertainties, owing to their narrowness and weakness 
and to the uncertainty in subtracting starlight from the host galaxy
(see Dong X.\ et al.\ 2012 for details).
Second, it is susceptible to dust reddening, which is also
difficult to measure from either the AGN continuum or the Balmer decrement
of the broad emission lines due to their faintness.
Although the measurements of the Balmer decrement 
are considered to be reliable for the D12 sample in a statistical sense,
which is consistent with no dust reddening for the bulk of the sample, 
the exact value for individual objects may suffer from large uncertainties. 

X-ray observation is a powerful tool to verify their AGN nature and to
 estimate the bolometric luminosity independently. 
Given their expected low luminosities,
not much brighter than off-nucleus ultra-luminous X-ray sources (ULXs),
both high spatial resolution and high sensitivity are needed.
As a pilot study along this line, 
 with Chandra, we observed a small sample of four candidate low-mass, 
low-\redd\ ($\sim10^{-2}$)   AGNs selected from the SDSS,
which are among the lowest \redd\ found of this class.
Among them,  two objects have 
significance levels among the lowest of SDSS low-mass AGNs
in terms of the detection of the  broad \hb\ and \ha\ lines,
owing partly to their weakness and narrowness 
and partly to the contaminating stellar spectrum.
In order to scrutinize the reliability of their broad \ha/\hb\ lines and, 
thus, their low-mass BH nature,
we also observed these two objects with 
the Multi-Mirror Telescope (MMT) to acquire
optical spectra with better spatial resolution 
(i.e. smaller aperture size, so as to reduce starlight contamination)
and higher signal-to-noise ratio (S/N) than those of the SDSS spectra.
In this paper, we report the results 
from our Chandra and MMT observations.
We assume a cosmology with $H_{0}$= 70 km\, s$^{-1}$\,Mpc$^{-1}$,
$\Omega_{M}=0.3$, and $\Omega_{\Lambda}=0.7$.

\section{Sample} 

\begin{figure}[]
\begin{minipage}[]{0.5\hsize}
\includegraphics[width=\hsize,angle=0]{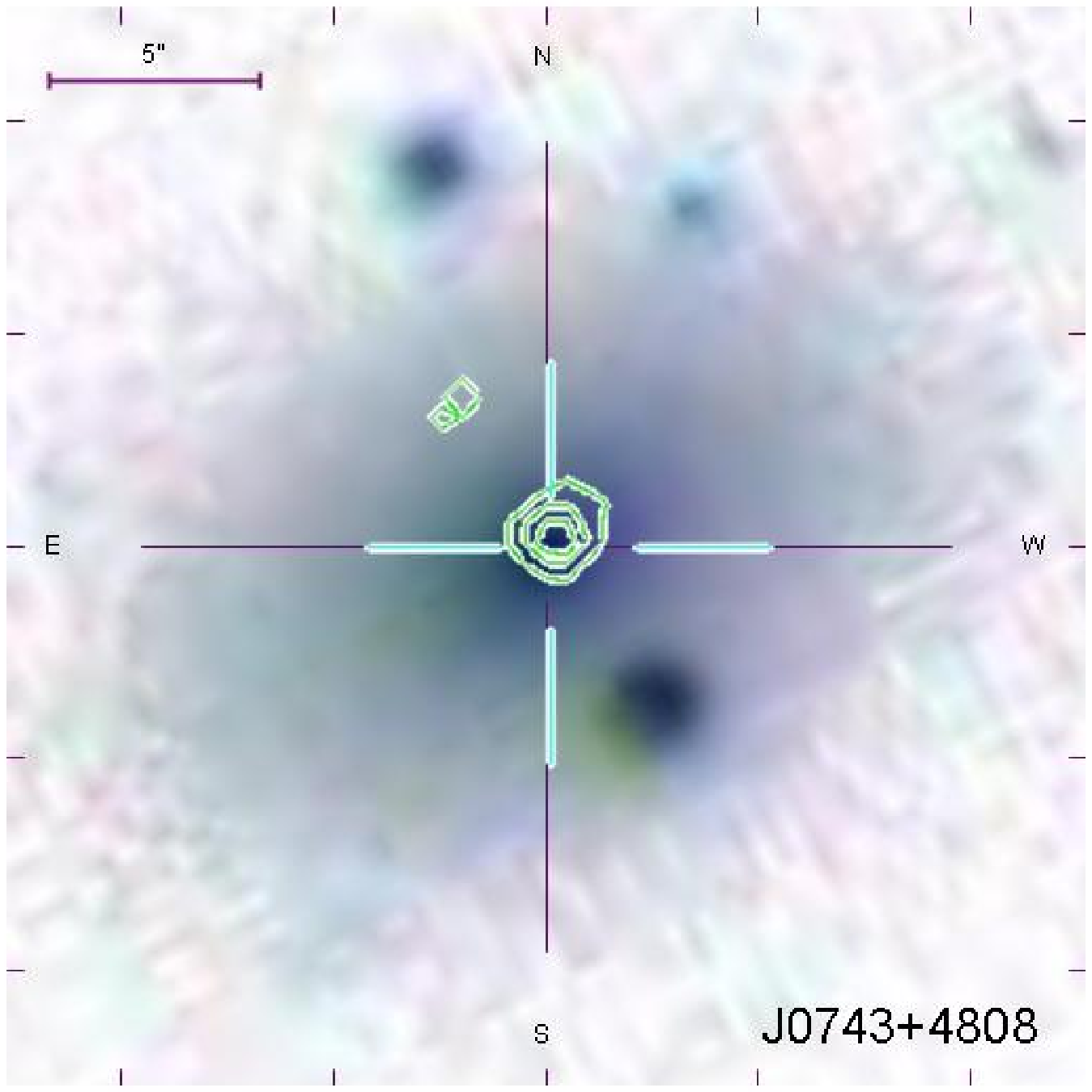}
\includegraphics[width=\hsize,angle=0]{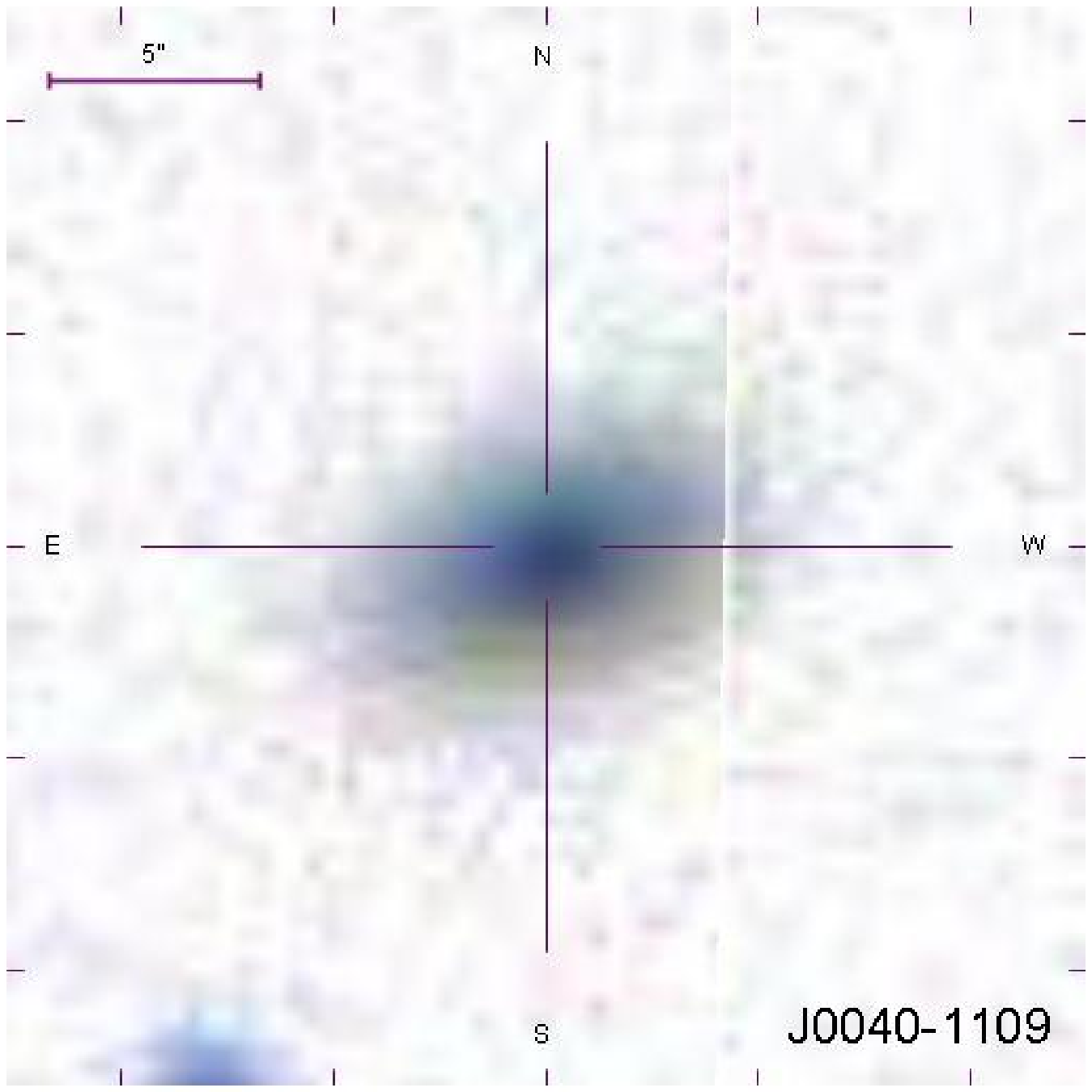}
\end{minipage}
\begin{minipage}[]{0.5\hsize}
\includegraphics[width=\hsize,angle=0]{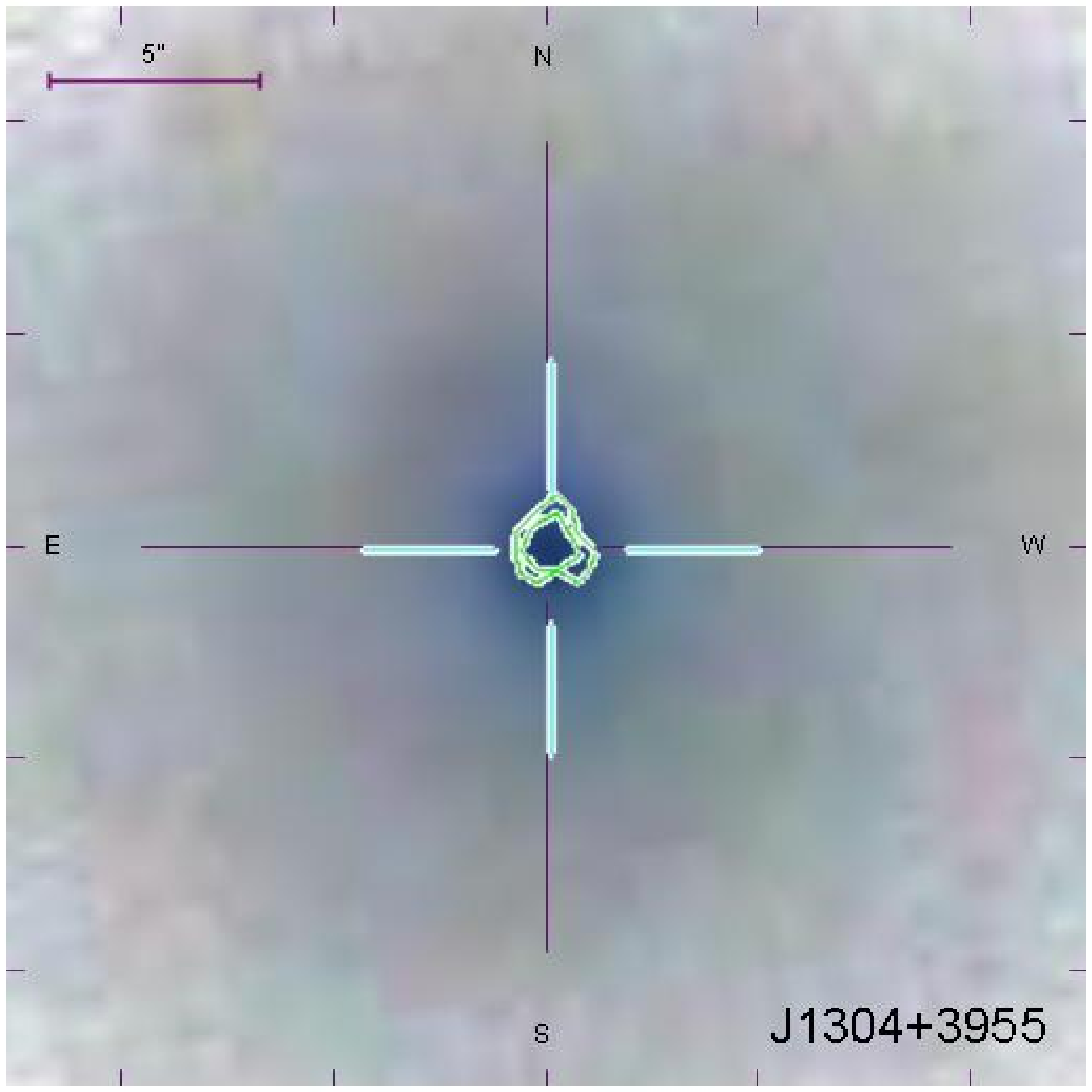}
\includegraphics[width=\hsize,angle=0]{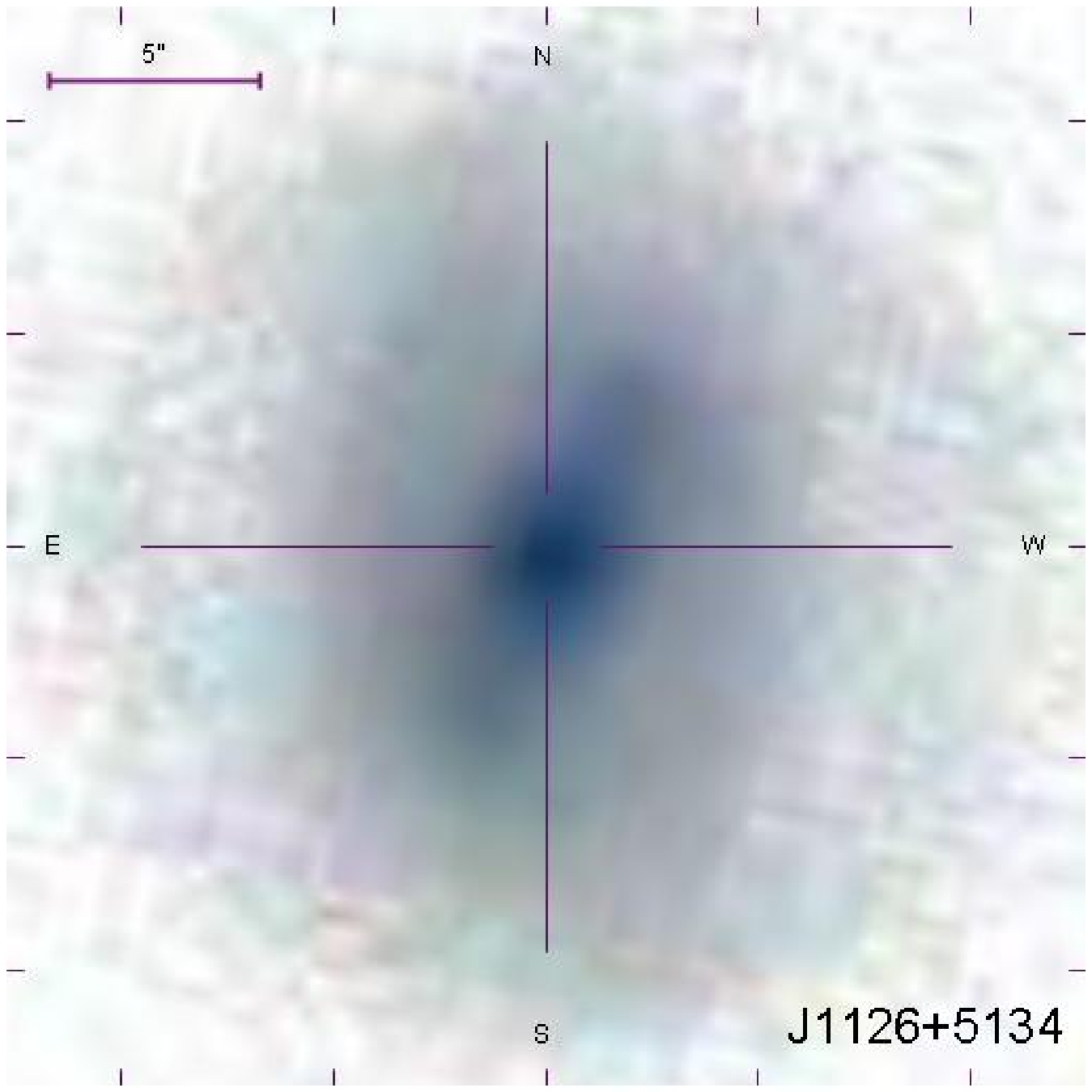}
\end{minipage}
\caption{
Optical images of the sample objects taken from the SDSS
(upper-left: J0743+4808; 
upper-right: J1304+3955;
lower-left: J0040$-$1109; and
lower-right: J1126+5134);
superimposed are the intensity contours of the X-ray images observed with Chandra
for the X-ray detections. 
The positions of the optical nucleus and the X-ray sources are indicated by
the purple and green crosses, respectively.  
The off-nucleus X-ray source, which is a ULX, 
detected in the galaxy J0743+4808
(north-east to the nucleus) is clearly seen.
}
\label{fig:opt_image}
\end{figure}

\begin{figure}
\epsscale{0.8}
\plotone{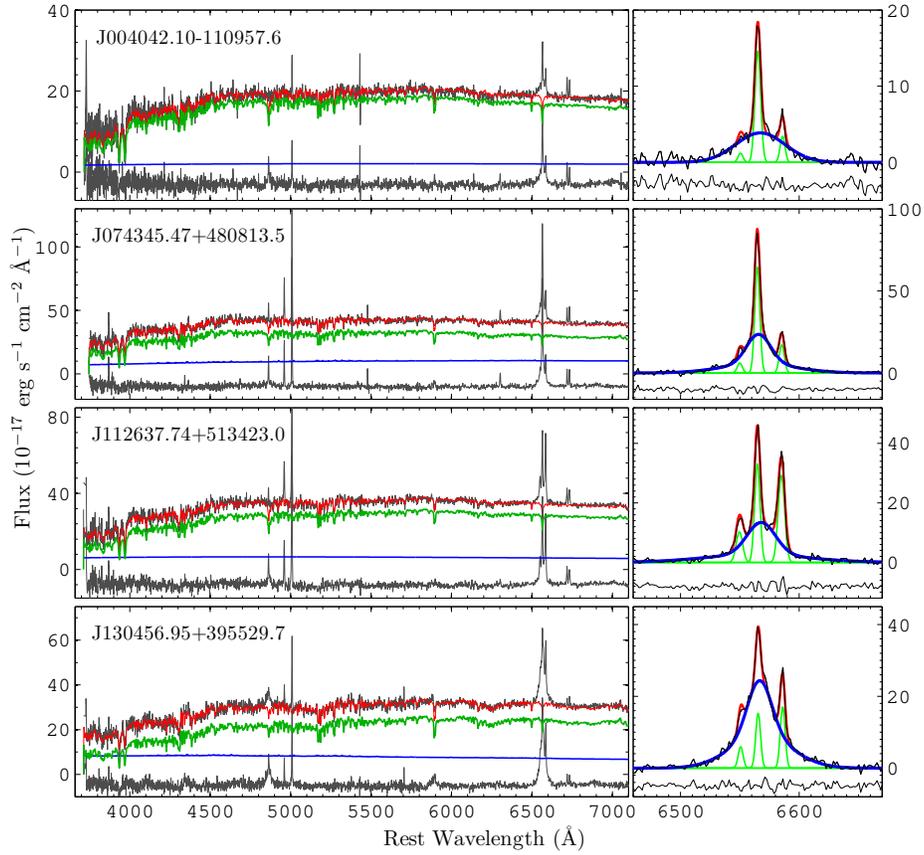}
\caption{Optical SDSS spectra and the best-fit models of the sample objects.
{\it Left panels}: the observed SDSS spectrum (black), the total model
(red), the decomposed components of the host galaxy (green) and the AGN
continuum (blue).
The starlight and continuum subtracted emission-line spectra are 
overplotted at the bottom. 
{\it Right panels}: emission-line spectra and model fitting in the \ha~$+$~\nii\ region.
 } 
\label{fig:sdssspec}
\end{figure}

We selected candidate low-accretion  low-mass AGNs from the SDSS DR5, and
a number of objects with the lowest Eddington ratios ($\sim 10^{-2}$) were found.
We chose the nearest four objects within $\sim$100\,Mpc
in consideration of easy detection in X-ray 
(Table\,\ref{tab:sample}).
Their optical images taken by the SDSS are shown in Figure\,\ref{fig:opt_image}.
Two objects among them,  
J004042.10$-$110957.6 and J112637.74+513423.0 
(referred to as J0040$-$1109 and J1126+5134 hereafter),
have been presented in the D12 sample selected from the SDSS DR4.
The remaining two objects, 
SDSS\,J074345.47+480813.5 (J0743+4808) 
and J130456.95+395529.7 (J1304+3955),
only available in the DR5, are selected  using exactly the same data analysis 
and selection procedures
\footnote{
We used an elaborate spectral analysis algorithm, which performs 
spectral decomposition of AGN and the host galaxy starlight and de-blending
of the broad and narrow emission lines. 
The black hole masses, bolometric luminosities and Eddington ratios 
were estimated in the same way as in \citet{gh07b},
which makes use of the luminosity and FWHM of the broad \ha\ line,
$L_{\rm bol} = 9.8 \lambda L_{\lambda} (5100$\,\AA) \citep{md04},
and the $\lambda5100$\,\AA--\ha\ luminosity relation \citep{gh05b}. 
See \citet{dong12} for details.
}
as described in \citet{dong12}. 
The spectra where they were identified,  
as well as those of the two objects in the D12 sample, 
are shown in Figure\,\ref{fig:sdssspec} 
along with the best-fit spectral models 
(see Dong X.-B.\ et al.\ 2012 for the method of spectral fitting).
The derived parameters are listed in Table\,\ref{tab:sample};
for the two in the D12 sample, the parameters are taken from that paper.
They have BH masses $\sim 10^6$\,\msun\ and the estimated Eddington ratios 
around a few percent.

As can be seen from their SDSS spectra in Figure\,\ref{fig:sdssspec}, 
a broad \ha\ component is evident in both J0040$-$1109  and J1304+3955;
a broad \hb\ component is also clearly present in J1304+3955, whereas
in J0040$-$1109 such a component can only be seen after the proper removal
of starlight.
However, these lines appear to be comparatively narrower and weaker
in  J0743+4808 and J1126+5134,
causing their detection significance not to be  high, 
but merely above the threshold adopted in \citet{dong12}.
Furthermore, the presence of a broad \hb\ line is elusive in these two objects.
To verify the detection of the broad \ha\ and possibly the broad \hb\ lines,
and thus their low-mass AGN nature,
we also observed these two objects with MMT to acquire
optical spectra with better spatial resolution and S/N.

\section{Results of MMT observations}

To scrutinize the reliability of the broad lines in J0743+4808 and J1126+5134,
we observed them with the blue-channel spectrograph 
of the 6.5m MMT telescope on 2008 February.
We used the setting of 500 l\,mm$^{-1}$ gratings blazed at 6000\AA, which
covers a wavelength range from the \hb\ to \ha\ lines.
A slit width of 1\arcsec\ was chosen to match the seeing;
this is essential for reducing starlight contamination.
The total exposure time was 30\,minutes for each object
(two frames taken, each with a 15 minute duration).
The corresponding spectral resolutions are 3.8\AA\ 
in FWHM,
as measured from the comparison lamp lines. 
A KPNO standard star was observed for flux calibration.
The data reductions, 
including bias subtraction, flat-field correction, and cosmic-ray removal, 
were accomplished with standard procedures using IRAF. 
One-dimensional spectra were extracted, 
and were calibrated using the observed standard star.  
Since the continua of both objects are dominated by starlight, 
only the (broad) emission lines can be used as an indicator of the AGN luminosity.
To improve the spectrophotometric calibration, 
we further recalibrate the resulting spectra using the [OIII] emission line 
against that in the corresponding SDSS spectra\footnote{
At a redshift of 0.018 (for J0743+4808; it is even higher for J1126+5134), 
1\,
arcsec corresponds to 360\,pc, 
which is much larger than the expected size of the narrow-line region (NLR).
Both the NLR and broad-line region (BLR) 
can thus be considered as a point source for the ground-based (MMT) observations. 
On the other hand, the narrow-line intensity ratios 
are almost the same for the MMT and SDSS spectra,
which are typical of Seyfert\,2 AGNs;
this indicates that the contamination of HII regions in the host galaxy is negligible.
Therefore, we could use the SDSS spectra for spectrophotometric calibration,
although they were taken within a fiber aperture of 3\arcsec, 
which is larger than the slit width used in our MMT observations.
}, 
which are generally
well calibrated to an accuracy of $\sim8$\%.
Galactic extinction is corrected by using the 
extinction map of \citet{sfd98}
and the reddening curve of \citet{fit99}. 
The redshifts measured from the SDSS spectra are used.

\begin{figure}
\epsscale{0.8}
\plotone{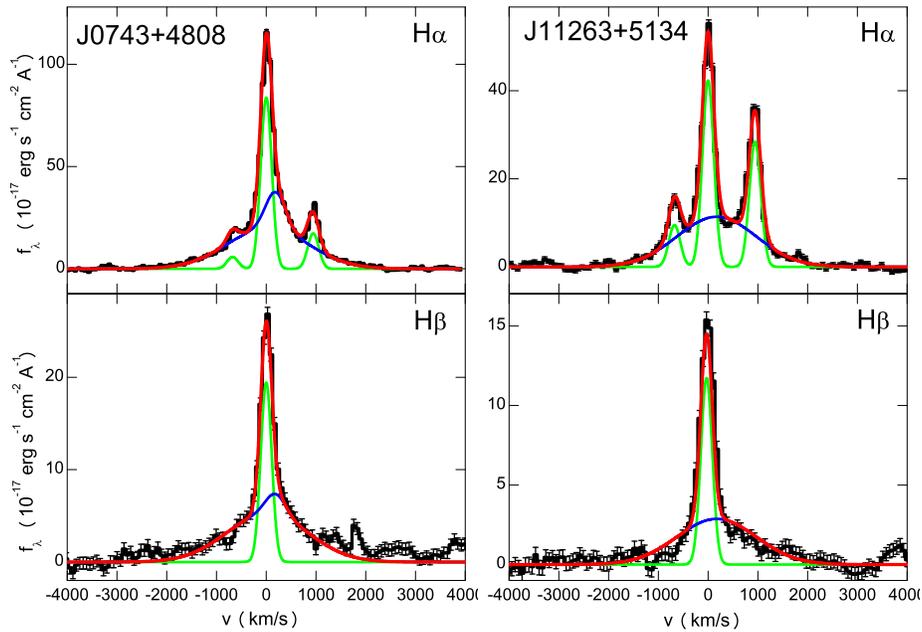}
\caption{
Emission line spectra (with the starlight and continuum subtracted)
 taken at the MMT telescope for J0743+4808 (left panel)
and J1126+5134 (right panel) 
in the \ha\  and \hb\ line regions,
as well as the best-fit models. 
Colour code: data (black); model: total (red), broad line (blue) and
narrow line (green).
} 
\label{fig:mmtspec}
\end{figure}

The spectral analysis is performed in the same way as that in D12
(see also Zhou et al.\ 2006), and is only outlined here.
Host galaxy starlight, 
though with considerably less contribution compared to the SDSS 
spectra---thanks to the 1\arcsec\ slit, 
is carefully modeled and removed using our algorithm 
described in \citet{lu06}. 
The emission lines are de-blended into a narrow and a broad component
by fitting multi-Gaussian component models, with the least number of
components required (set by a $F$-test).
As demonstrated in D12, these procedures work well for modeling
the optical spectra of low-mass AGNs.
The emission line spectra in the \ha\ and \hb\ regions 
of the reduced MMT spectra are shown in Figure\,\ref{fig:mmtspec}, 
along with the best-fit emission line models.

The MMT spectra have much higher S/N than the SDSS spectra.
In both objects, broad components of both \ha\ and \hb\ 
are clearly detected.
The measured line widths, with smaller errors compared to
the SDSS values, are generally consistent with the SDSS measurements
within typical mutual uncertainties.
We derive the BH masses and Eddington ratios using 
the new MMT measurements of the \ha\ line width,\footnote{
Note that we use the broad \ha\ luminosities of the SDSS spectra in this calculation,
in consideration of the light loss due to the small slit aperture
used in the MMT observations.} as given in Table\,\ref{tab:sample}.
The new \mbh\ and \redd\ values are within a factor of $1.5$
compared to those from the SDSS, which are well within the
statistical uncertainties of the BH masses estimated from the SDSS data
\citep[0.3\,dex, see][]{dong12}. 
We use the new \mbh\ and \redd\ values for these two objects 
in the rest of the paper.
To conclude,  J0743+4808 and J1126+5134 are confirmed to be
type\,1 AGNs with a low-mass BH and a relatively low, \ha-derived Eddington ratio.
This suggests that the low-mass AGNs with low \redd\ 
found in the D12 sample should largely be reliable.

\section{Results of Chandra X-ray observations}

\subsection{Data Reduction and Source Detection}

We performed snapshot observations for 
the four low-mass BHs with Chandra ACIS-3, each for 5\,ks,
as a Guest Observer program (observation ID 09233-09236).
The observation logs are listed in Table\,\ref{tab:xobs}.
The data analysis is performed using the standard CIAO tools. 
A level 2 event file is created from 
the provided level 1 data following the standard procedure.
Events with grades of 0, 2, 3, 4 and 6 in the 0.3--8\,keV  
energy range are selected. 
Source detection is performed using the "celldetect" task 
with default parameters. 
Of the four objects, J0743+4808 and J1304+3955 are significantly detected
at the $\sim$14$\,\sigma$ and $\sim$10$\,\sigma$ significance levels, respectively.
The X-ray sources coincide with the positions of the  galactic nuclei in optical
given by the SDSS. 
We extract their source X-ray counts using a 3\arcsec\ radius aperture,
and the background counts using an annulus of inner and outer radii of
6\arcsec\ and 9\arcsec, respectively.
J0743+4808 and J1304+3955 have net source counts of 
194 and 94, respectively.
The observational logs and source detection results are summarized 
in Table\,\ref{tab:xobs}.

For the remaining two objects not detected by the source detection algorithm, 
X-ray counts are extracted from a 2\arcsec\ radius aperture 
centered at the optical positions of the nuclei,
yielding 2 counts for J1126+5134 and 2.5 counts for J0040-1109.
The expected local background counts within the aperture are estimated by averaging
the background in a much larger source-free region around the optical position,
yielding  0.16 and 0.27 for J1126+5134 and J0040-1109, respectively.
The probability of obtaining the counts as observed or more when the above mean 
background counts are expected is $1\times 10^{-2}$ for J1126+5134
and  $3\times10^{-3}$ for J0040-1109, based on Poisson statistics.
This may be a hint for the possible existence of an X-ray source, as discussed in 
 \citet{vig01}.
However, detections  at such significance levels are uncertain or  marginal at best,
and thus we  do not consider them as reliable detections.

Here we estimate the upper bound of the X-ray counts (mean $\bar{s}$ expected)
from a postulated X-ray source at the optical position within an exposure time $T$,  
making use of both the 
observed counts $N$ and the estimated mean of the background ($\bar{b}$).
The total mean  counts  expected are $m=\bar{s}+ \bar{b}$.
The Poisson probability of obtaining  $N+1$ counts or more when $m$ counts are expected
is $p=P(\ge N+1\mid m)$. As a conservative case, 
we require this probability to be as high as $p=90\%$ or, 
equivalently,  $P(\le N \mid m)=0.1$.
For given observed counts $N$, the mean counts $m$ can be obtained, 
which we consider as the upper limit at the $90\%$ confidence level, 
since  for any emission with a  mean smaller than $m$ only lower counts ($\le N$) 
could be obtained at the same probability.
In this way, we find $m$=5.32 and 6.68 
for J1126+5134 ($N=2$) and  J0040-1109 ($N\approx3$), respectively.
Thus, the upper bound of the expected counts of any possible source can be obtained as 
$\bar{s}=m- \bar{b}$.
Adopting the $\bar{b}$ values  derived above, we find 
$\bar{s}=5.16$ for J1126+5134 and $6.41$ for J0040-1109.   
The upper bounds on the expected count rates are obtained as 
$1.1\times10^{-3}$\,\ucr\ and $1.4\times10^{-3}$\,\ucr\ 
for J1126+5134 and  J0040-1109, respectively, in 0.3--8\,keV.

\subsection{X-Ray Spectra and Luminosities}

The detected X-ray counts of J0743+4808 and J1304+3955 are just 
enough for a rough characterization of their X-ray spectral properties.
This is very useful for determining the intrinsic X-ray luminosities 
(and the Eddington ratios) of the objects, which is the main goal of this study.
The X-ray spectra of the source and backgrounds are extracted, respectively, from 
the above regions in the 0.3--8\,keV band. 
The RMF and ARF files are created at the source positions using CIAO. 
We use XSPEC for spectral fitting. 
Galactic absorption is always included and the HI column density \nh\ is 
fixed in the spectral fit.
The uncertainties of  the parameters derived from the 
X-ray data are quoted at the 68\% confidence level throughout the paper.

\subsubsection{J0743+4808}

\begin{figure}
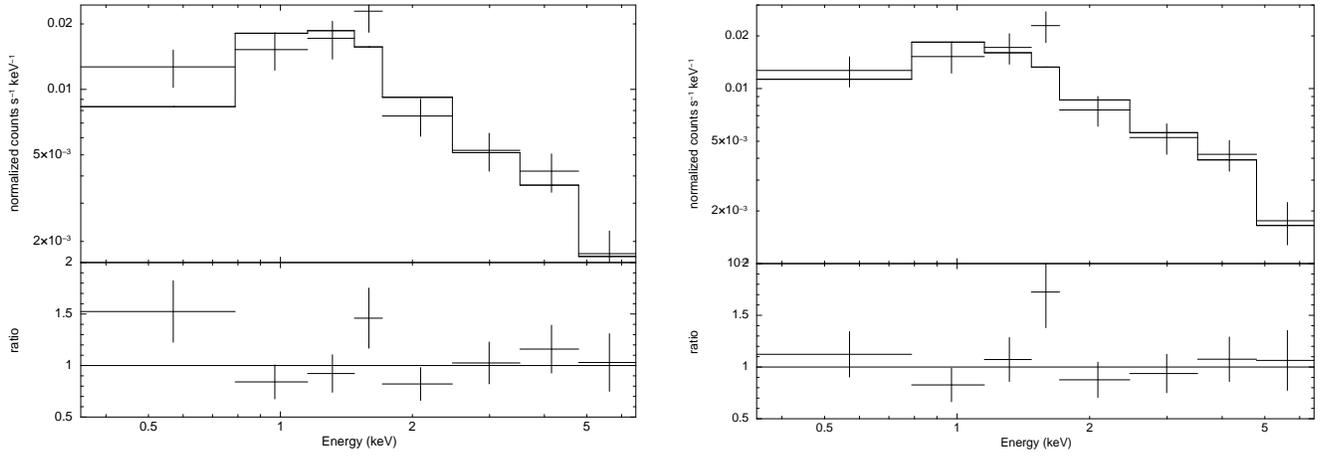

\begin{minipage}[]{0.47\hsize}
\includegraphics[width=0.7\hsize,angle=-90]{plot_fit_pl-j07.eps}
\end{minipage}
\hfill
\begin{minipage}[]{0.47\hsize}
\includegraphics[width=0.7\hsize,angle=-90]{plot_fit_zpcfabs-j07.eps}
\end{minipage}
\caption{
{\bf Chandra X-ray spectra and the best-fit models for 
J0743+4808; left: a flat power-law with Galactic absorption;
right: a partial covering 'warm' absorption model 
 (see text)}.
} 
\label{fig:xspec}
\end{figure}

The spectrum of J0743+4808  is binned to have $\sim$25 net source counts 
in each bin to achieve a S/N$\sim 5$, 
resulting in eight energy bins (Figure\,\ref{fig:xspec}).
The fitting results are given in Table\,\ref{tab:xfit}.
A power-law model with Galactic absorption 
gives a statistically acceptable fit (\pnull=0.21), 
resulting in a  photon index $\Gamma=1.02\pm0.13$ and a 2--10\,keV luminosity
$L_{\rm 2-10keV}=3.7\times10^{41}$\,\ulum.
The spectrum and the best-fit power-law model are shown 
in Figure\,\ref{fig:xspec} (left-hand panel).
Setting the (neutral) absorption column \nh\ as a free parameter leads to
an even flatter spectrum ($\Gamma<1$) and, in the worst case, 
no absorption occurring (cf. the Galactic value $5.4\times 10^{20}$\,\unh), 
which is physically unacceptable. 
Setting $\Gamma$ to steeper values, such as the typical value for Seyfert\,1 galaxies
$\Gamma$=1.7 \citep{mus93} in 2--10\,keV,  
 yields no acceptable fits  
 (reduced \chisq=4.1 for $\Gamma$=1.7 and \nh\ fixed at the Galactic value).
We thus conclude that the observed X-ray spectrum is apparently flat. 
Interestingly, a similar flat X-ray spectrum was also observed occasionally in NGC\,4395, 
which resembles J0743+4808 in  both \mbh\ and \redd.
However,  it is well known that an apparently flat X-ray continuum may result from reflection
or absorption (ionized and/or partial covering) of an intrinsically steep continuum.
Here we also explore these alternative models.

Reflection models, either neutral or ionized 
({\it pexrav} and {\it pexriv} in XSPEC, respectively), 
yield unacceptable or at most marginally acceptable fits over a 
range of $\Gamma$ values (from $\Gamma=1.0$ to 2.0) and, in the worst case,
require an unphysically large reflection albedo ($\gg 1$).
Thus they are not considered to be viable models.

Here we consider the case where the power-law continuum is absorbed
and adopt the ionized absorption model {\it zxipcf}, 
that is parameterized by a column density \nh, an ionization parameter $\xi$\footnote{
Defined as $\xi=L/nr^2$, where $L$ is the ionizing luminosity,
$n$ the ion density and $r$ the distance of the absorber to the 
central ionizing source.
} 
and a covering factor $f_{\rm c}$.
First, when fixing the covering factor to unity,
the best fit is achieved with an ionized absorber ($\log\xi\sim2.4$) and
a steeper continuum than above ($\Gamma=1.31^{+0.21}_{-0.24}$),
resulting in $\chi^2$=4.4 for 4 degrees of freedom (dof.), \pnull=0.35.
The best-fit model is shown in Figure\,\ref{fig:xspec} (right-hand  panel).
Freely fitting the covering factor also leads to $f_{\rm c}=1$.
The fitted power-law photon index is somewhat flatter than,
but still marginally consistent with (within 90\% uncertainty), 
the typical value for Seyfert galaxies, $\Gamma= 1.7$. 
Alternatively,
we further assume $\Gamma= 1.7$ for the intrinsic continuum.
Fixing $f_{\rm c}=1$ yields an acceptable fit (\pnull=0.24),
with a moderately ionized absorber of \nh$\sim2 \times 10^{22}$\,\unh.
Fitting $f_{\rm c}$ as a free parameter
improves the fit only marginally (\pnull=0.26), 
and results in $f_{\rm c}=0.78^{+0.10}_{-0.09}$ and a slightly increased \nh. 
We also try an absorber model with low ionization by fixing $\xi$ 
at the lower bound $\log\xi=-3$, 
which gives similar fitting statistics
(\pnull=0.26) and covering factor, but reduced \nh;
in this case a partial covering absorber is required, 
as full coverage can be ruled out (\pnull=0.01).
In summary, the observed spectrum of J0743+4808 can also be reproduced 
by a steep  power-law ($\Gamma= 1.3-1.7$) subject to absorption with 
\nh\ of the order of $10^{22}$\,\unh;
the ionization status is poorly constrained, ranging from
"cold" absorber with a covering factor $\sim0.7$ 
to ionized gas with a large or full coverage.

All the acceptable models above give similar flux densities and luminosities
(corrected for intrinsic absorption) in the 2--10\,keV band, 
which are actually consistent with one another
within mutual statistical errors (Table\,\ref{tab:xfit}).
In the analysis below, we take the representative value 
$L_{\rm 2-10keV}=4.4(\pm1.5)\times10^{41}$\,\ulum\
as the nominal luminosity for the intrinsic absorption model,
which is the case for fixing  $\Gamma= 1.7$ and freely fitting 
{\em all} the absorption parameters;
its  uncertainty is assigned as
either the statistical errors or the systematic luminosity differences 
among the various models above, whichever is the largest.

\subsubsection{J1304+3955}

Given the low source counts, we bin the spectrum to have 
at least 15 counts per energy bin,
and adopt the C-statistic in the spectral fitting.
A power-law with Galactic absorption yields an excellent fit
(C-statistic = 1.5 for six data bins),
and a photon index $\Gamma=1.70\pm0.18$, which is well consistent
with the typical values of Seyfert galaxies.
We also try the absorption models as in J0743+4808, and find that
the fitted absorption \nh\ is consistent with zero, 
leading to  almost the same fitting results.
Thus the X-ray spectrum of J1304+3955 can be described by a simple power-law 
(with Galactic absorption)
with an index typical of normal Seyfert galaxies. 
The spectrum and the best-fit power-law model are shown 
in Figure\,\ref{fig:xspec}.
The luminosity in the 2--10\,keV band is 
$L_{\rm 2-10keV}=1.84\times10^{41}$\,\ulum.

\subsubsection{X-Ray Non-detections}
\label{noxray}

For the two undetected objects, upper bounds on their 
X-ray luminosities are derived using XSPEC as the following.
First a spectral model (absorbed by the Galactic \nh) 
is assumed with the normalization set to an arbitrary value, 
and is convolved with the response matrix and effective area (the RMF and ARF files).
Then the  normalization is determined by matching 
the predicted count rates in the 0.3--8\,keV band
to the upper bounds derived from the observations above.
Assuming a power-law model with two distinctive  photon indices,
$\Gamma=1.7$ as found in J1304+3955 (and in typical Seyfert galaxies) 
and $\Gamma=1.0$ in J0743+4808,
we find, for $\Gamma=1.7(1.0)$,  
$L_{\rm 2-10keV}<0.97 (2.1) \times10^{40}$\,\ulum\  for J1126+5134 and
$L_{\rm 2-10keV}<1.3 (2.8) \times10^{40}$\,\ulum\  for J0040-1109
 (see Table\,\ref{tab:xlum}).
Note that assuming the alternative intrinsic absorption model found in J0743+4808
yields  similar luminosity limits.
 The X-ray luminosities are constrained to be extremely low, 
 comparable to or even lower than some ULXs.

\subsection{X-Ray Variability}

\begin{figure}
\plottwo{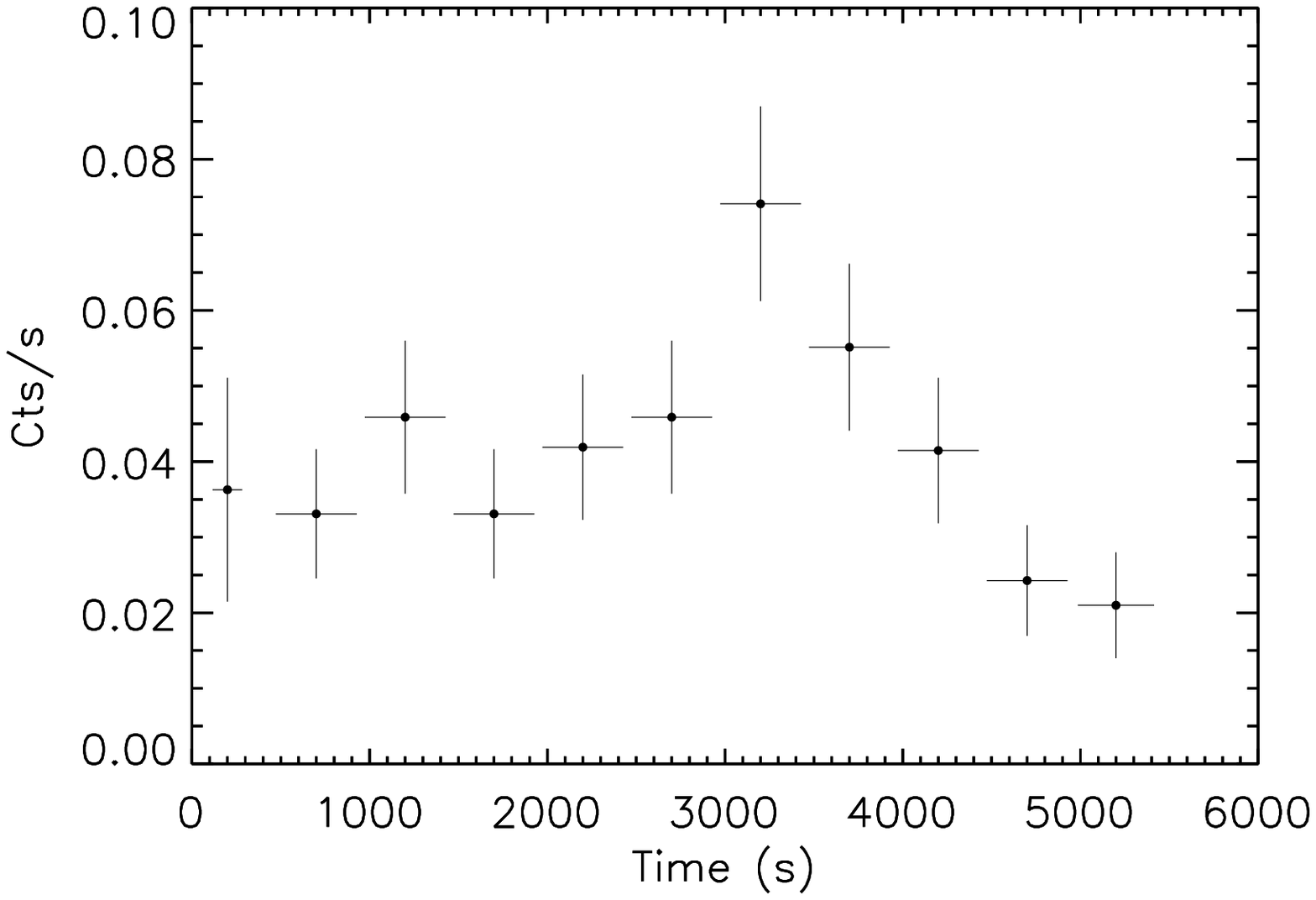}{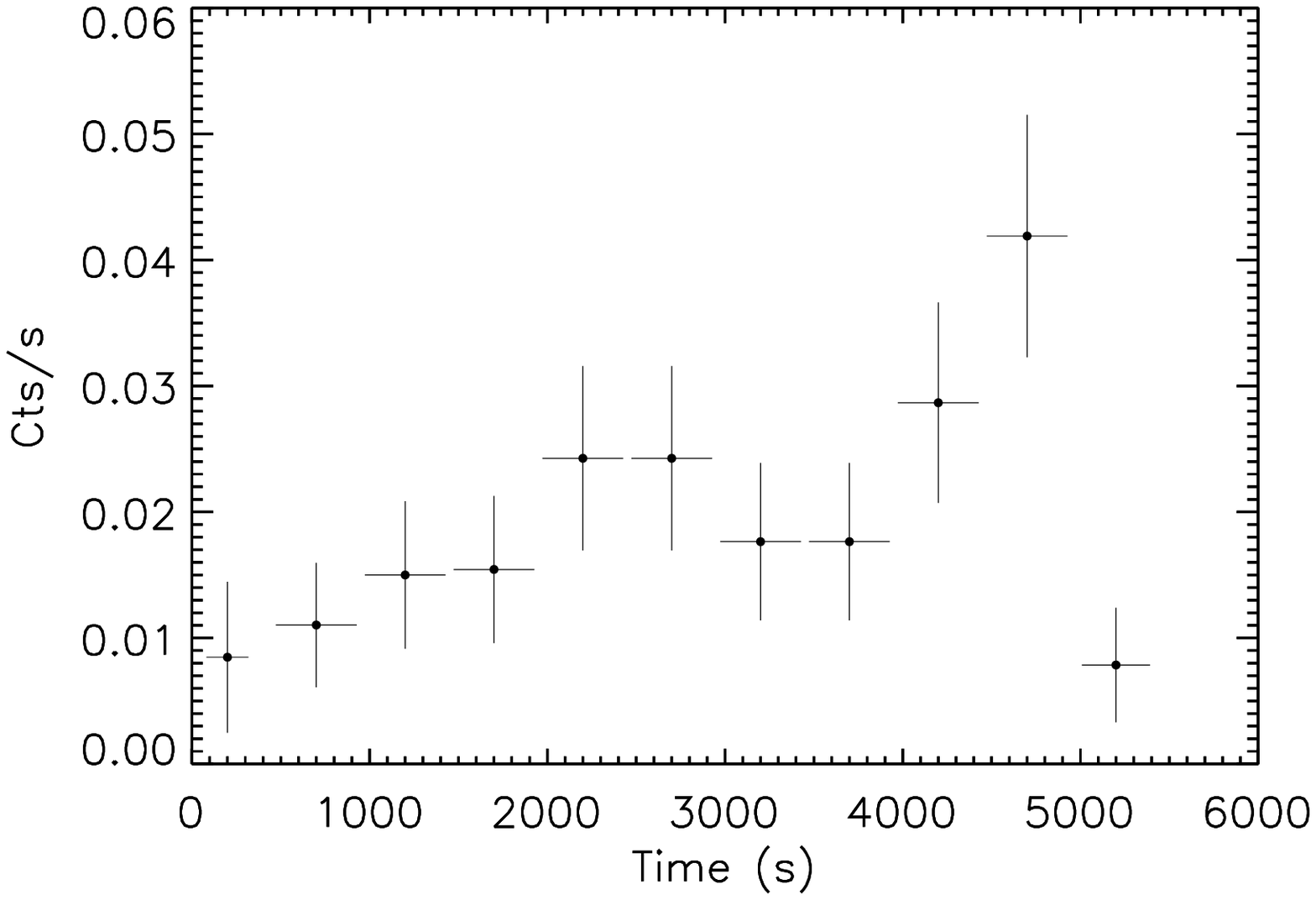}
\caption{
Background-subtracted X-ray lightcurves of J0743+4808 (left) and J1304+3955 (right); the bin sizes are 500\,seconds (except for the first bin).
} 
\label{fig:xlc}
\end{figure}

For J0743+4808 and J1304+3955 background-subtracted lightcurves 
in 0.3--8\,keV are constructed, using bin sizes of 500\,s
for all except one of the bins (Figure\,\ref{fig:xlc}).
As can be seen,  for both objects,
the X-ray count rates appear to be variable within 
the observational intervals.
We test hypothesized constant X-ray counts 
over time bins of 500\,seconds, using the $\chi^2$ test \citep{bev92}.
This yields a null probability \pnull=0.9\% and \pnull=4.6\% for 
J0743+4808 and J1304+3955, respectively, 
indicating at least marginally significant variability.
The proper determination of the variability timescales is limited by
the modest count rates and exposure time, however. 
Here we estimate the timescales as the intervals between the 
minimum and maximum fluxes which varied by at least a factor of two.
This results in  about 2000\,s or less in both objects.
Such timescales are short for AGNs, and are consistent with 
what has been observed in NGC\,4395 and 
Pox\,52 \citep{mor05,vaug05,dew08}, and other low-mass AGNs 
\citep[e.g.][]{dew08,min09,ai11}.

\subsection{Off-nucleus X-Ray Source in J0743+4808}
\label{sect:ulx}
Interestingly, in the galaxy J0743+4808, 
an off-nucleus X-ray source (denoted as J0743+4808\,X-1), 
is detected at 3.3$\,\sigma$ significance, 
with 12 source counts and a count rate of $2.32\times10^{-3}$\,\ucr.
Its position (RA=07h43m45.7s, Dec=+43d08m17s) 
is 4.\arcsec3 away from the nucleus, 
corresponding to a projected physical distance of 1.62\,kpc 
(Figure\,\ref{fig:opt_image}).
We calculate its luminosity by deriving 
the count rate to flux conversion factor,
assuming a power-law spectrum without intrinsic absorption.
Assuming $\Gamma=2.0 (1.8)$, 
we obtain a flux density of $8.2(12.2)\times10^{-15}$\,\ergs\ and a luminosity
$6.0 (8.9) \times10^{39}$\,\ulum\ in 2-10\,keV.
Clearly, it is a new ULX detected in dwarf galaxies.
Assuming that its X-ray emission is not beamed, 
the observed X-ray luminosity sets a lower limit on the 
BH mass of $\sim$48(70)\,\msun\ 
(if the BH is accreting at a rate not significantly higher than the Eddington rate,
as commonly believed).
No optical counterpart is found based on the SDSS image.

\section{Estimation of Eddington ratios}
\label{sect:er}
The ratios of the 2--10\,keV luminosities to the Eddington luminosities are calculated
(Table\,\ref{tab:xlum}).
They are a few times $10^{-3}$ for the two X-ray detected objects
(for J0743+4808,  the spectral modelings with or without 
intrinsic absorption lead to only small differences).
For the  two non-detections,
the upper limits are on  the order of $1\times 10^{-4}$, 
depending on the  power-law photon index assumed. 
To estimate the bolometric luminosities
 and the Eddington ratios, 
the hard X-ray to bolometric correction factor
$k=L_{\rm bol}/L_{\rm X}$ needs to be known.
The determination of $k$ is difficult, requiring 
simultaneous observations of the {\em nuclear} broad-band 
luminosities of a sample of AGNs. 
Albeit with large scatters, recent studies suggest that 
$k$ clusters around 10--20  
for typical Seyfert galaxies with \redd$\leq 0.1$
and \mbh\ mostly in the range $10^7-10^8$\,\msun\  
(Vasudevan et al.\ 2009, see also Vasudevan \& Fabian 2007, 2009; Ho 2009).
For low-mass AGNs, however, there is a lack of $k$ estimation, 
except for the intensively studied prototype NGC\,4395, which has  
a small correction $k=6.6$ \citep{mor05}.
As a reasonable assumption, we adopt the X-ray bolometric correction 
factor to be within the above broad range, i.e. $k=7-20$, 
for low-mass AGNs.

For J1304+3955, the Eddington ratio is
estimated to be \reddx$\simeq 0.01-0.03$, somewhat smaller than or at most
comparable to the \ha-based estimation.
For J0743+4808, it is $0.03-0.1$; although
its lower end is comparable to the \ha-based estimation,
its upper bound is no longer in the low accretion regime.
For the two non-detections, the upper limits on \reddx\ 
are in the range \reddx$\la (0.4-3.4)\times10^{-3}$,
depending on $k$ and  $\Gamma$ assumed.
If this is the case, 
J1126+5134 and  J0040-1109 would be intrinsically extremely weak in X-rays
and also highly sub-Eddington, 
and the upper bounds are comparable to that of NGC\,4395 
\citep[][see, however, Nardini \& Risaliti (2011) for a different estimate of  ~0.01]{pet05}.

We note that the Eddington ratios constrained for the two non-detections 
are at least one order of magnitude lower compared to their values derived from the \ha\ line.
This discrepancy may be accounted for by a few factors, or a combination of them.
First, variability of the radiation in  the X-ray or/and \ha, as often seen in AGNs,
can never be ruled out.
Second, low-mass AGNs may have different broad-band SEDs and consequently different 
bolometric corrections from massive Seyfert galaxies,
given the (albeit weak) dependence of the temperature of  accretion disks 
on BH mass (see also Section\,\ref{dis:prop} in below).
Third, the X-ray emission in these two objects may be attenuated by absorbers 
with a large covering factor.  

Here we investigate the effect of  X-ray absorption on the derived Eddington ratios.
X-ray absorbers are sometimes found in massive Seyfert galaxies, as well as in 
NGC\,4395 \citep[][ionised absorber with \nh$\sim 2\times 10^{22}$\,\unh]{iwa10} 
and potentially in  J0743+4808 studied in this work. 
The column densities of the absorbers are at most on the order of $10^{22}$\,\unh\
for  type\,1 Seyfert galaxies \citep[e.g.][]{mat10}.
We re-estimate the limits on the "intrinsic" X-ray luminosities of these two objects by
assuming the presence of absorbers with the condition similar to that inferred in J0743+4808
(we adopt the case of a partial covering,  ionized an absorber with 
\nh=$5.4\times10^{22}$\,\unh; see Table\,3), using the method described in Section\,\ref{noxray}.
For both objects, this results in upper limits on the Eddington ratio 
\reddx$<(2-4)\times 10^{-3}$ for a covering factor of 0.78 as in J0743+4808, 
which are similar to the case assuming a flat power-law of $\Gamma=1.0$.
In the extreme case where the covering factor is unity, 
we find \reddx$<(4-12)\times 10^{-3}$, and its
upper bound  becomes comparable to those derived from \ha.
We consider these values to be the most conservative limits for both objects. 

Thus the Chandra observations confirm the low Eddington ratios in most, if not all, of our objects, 
which are about a few percent or even lower,
for a reasonable range of the X-ray-to-bolometric correction factor.
These values are generally compared to, or even smaller than, 
the estimates from optical using \ha.
The true Eddington ratios may lie somewhere in between the values estimated in these two bands.

\section{Discussion}

\subsection{The Nature of an AGN with a Low-mass Black Hole}

The MMT optical spectra of J0743+4808 and J1126+5134
confirm the presence of a broad component in the Balmer lines 
suggested by the SDSS spectra, 
with the FWHM measurements consistent with those from the SDSS.
These results confirm their type\,1 AGN nature and BH mass estimates.
The two objects detected in X-ray with Chandra, J0743+4808 and J1304+3955,
both show X-ray spectra (see Section\,6.2 for discussion)
typical of Seyfert galaxies and
rapid variability on timescales as short as $\sim 2\times 10^3$\,s.
These properties indicate that the observed X-rays must originate from an
AGN,  rather than dominated by hot gas or integrated emission from
a population of stellar-mass X-ray sources in the nuclear region,
which might be a concern given the very low X-ray luminosities observed.
Their X-ray luminosities, on the order of
$10^{40-41}$\,\ulum\ or even lower,  are at the lowest luminosity end of 
Sefyert galaxies,  and actually  comparable to bright ULXs. 
A few Seyfert galaxies in this luminosity regime have been found 
in the literature \citep[e.g.][]{tho09}, including NGC\,4395.

The variability timescales on the order of $10^3$\,s
 are among the shortest observed in radio-quiet AGNs, and
comparable to what is observed 
in NGC\,4395 \citep[e.g.][]{iwa00,iwa10,mor05} and
other AGNs with small BH masses \citep{min09,ai11}. 
Specifically, NGC\,4395
shows a "break frequency" at $\sim 10^{-3}$\,Hz 
in the X-ray power spectrum density \citep{vaug05}, 
for a BH mass of 
$\sim4\times10^5$\,\msun\ \citep{pet05}.
These values are comparable to the dynamical timescale $t_{\rm dy}$---the 
shortest characteristic timescales associated with accretion flows---at 
small radii of the accretion disk, 
where the hot corona producing the hard X-rays is located.
We have 
$t_{\rm dy}=10^4 (R/3R_{\rm s})^{3/2} (M_{\rm bh}/10^8\msun$)\,s, 
where $R_{\rm s}$ is the Schwarzschild radius \citep{cze06};
for $M_{\rm bh}=10^6$\msun\ and an X-ray-emitting region at $R\sim10\,R_{\rm s}$,
$t_{\rm dy}\sim 600$\,s.
Such short time-scale variations are indicative of  
small BH masses ($\la10^6$\,\msun) in these objects.

\subsection{X-ray Spectral Property}

The two Chandra detected objects show distinct X-ray spectral shapes.
While J1304+3955 exhibits an unabsorbed power-law 
with a photon index ($\Gamma\sim1.70\pm0.18$) typical of Seyfert galaxies,
J0743+4808 shows an unusually flat spectrum ($\Gamma\sim1.02\pm0.13$).
Interestingly, a flat X-ray spectrum was also observed occasionally in NGC\,4395, 
which is similar to J0743+4808 in  both \mbh\ and \redd.
In fact, its 1--10\,keV spectrum is strongly variable from $\Gamma=1.7$ 
\citep{shi03,iwa10} to even $0.6$ \citep{mor05}
on timescales of a year or less, which is mainly accounted for 
by changes in the emission continuum shape with little changes in flux 
(rather than changes in the warm absorber).
If the observed flat spectra are intrinsic, 
the diverse spectral shapes in these two  objects as well as in NGC\,4395
indicate that, at the low \redd\ and \mbh\ regime, 
the physical conditions of the corona producing X-rays 
differ largely from one object to another and strongly vary
(the former may be a consequence of the latter).
Such a result seems not to comply with the claimed $\Gamma$--\redd\  relation\footnote
{A correlation has been suggested recently 
between $\Gamma$ and the Eddington ratio \citep[e.g.][]{ris09},
that predicts a flatter $\Gamma$ at lower \redd.
However, it is not clear whether this relation holds at the low \redd\ range studied here,
given the sparse sampling and large scatters 
(see their Figure\,2).
},
or is at least suggestive of a large scatter at the low-\redd\ end.
Spectral fitting of the X-ray data of a few small samples of low-mass AGNs,
taken with both XMM-Newton and Chandra, show $\Gamma$=1.5--2.7
\citep{dew08,min09,des09,ai11,dgh12};
however, all those objects are accreting at high rates around Eddington 
or a substantial fraction of Eddington.

Alternatively, as shown above, 
the flat spectrum found in J0743+4808 may also be reproduced 
with a steep power-law undergoing X-ray absorption, 
either ionized and/or partial covering, on the order of $10^{22}$\,\unh.
Partially ionized absorption is not uncommon in Seyfert\,1 galaxies, and so is 
a partially obscuring absorber (termed a partially obscured AGN).
The derived absorption parameters are also typical of those observed in 
type\,1 AGNs.
For low-mass AGNs,
partially ionized absorption has been detected in NGC\,4395
\citep{iwa00,shi03,mor05} and also possibly in Pox\,52 \citep{dew08}.
Future X-ray observations with better S/N will 
be able to distinguish the two cases---intrinsic 
flat spectrum or ionized/partial absorption.

\subsection{Properties of Low-mass AGNs Accreting at Low Rates}
\label{dis:prop}

\begin{figure}
\epsscale{1.1}
\plottwo{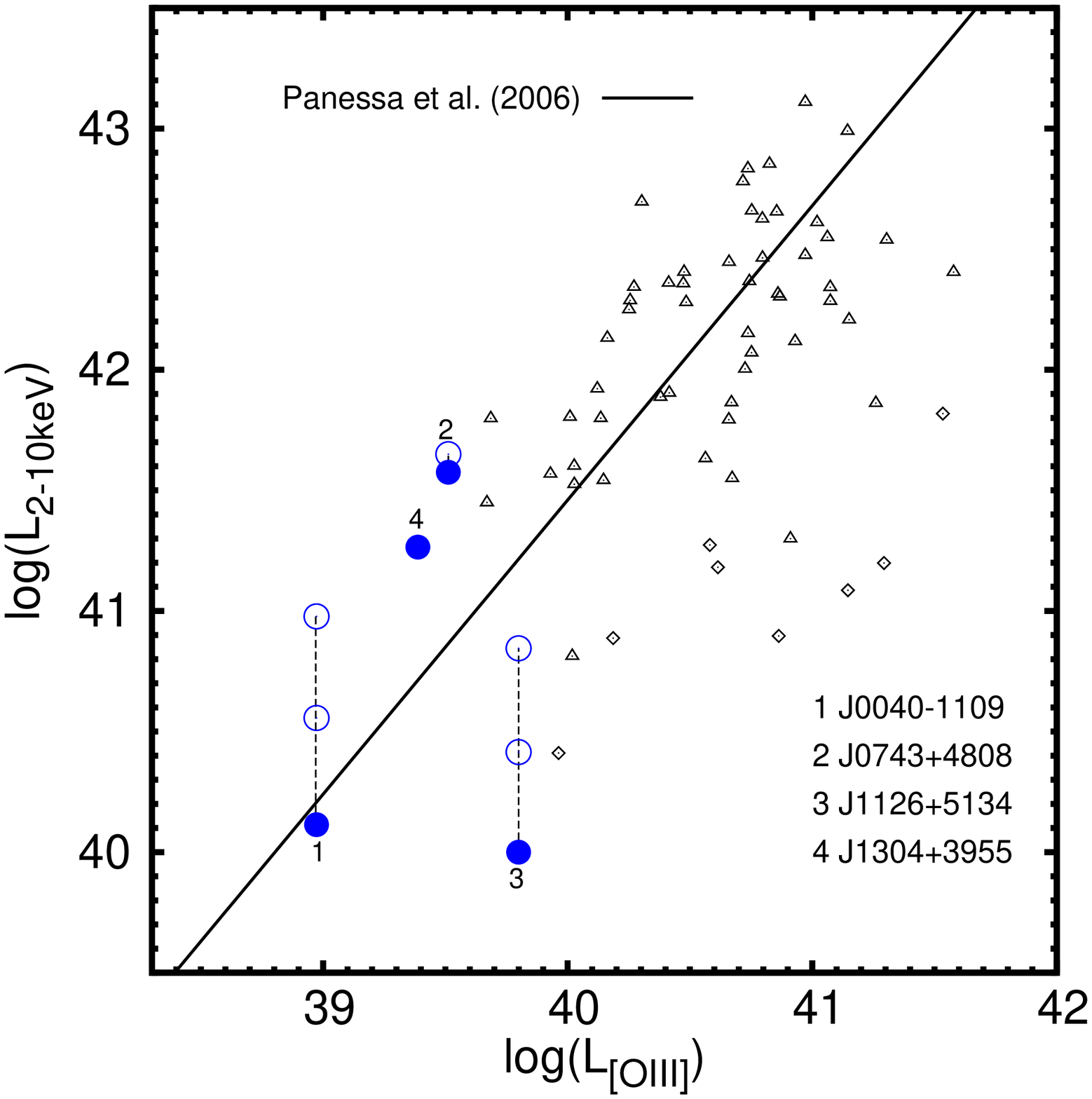}{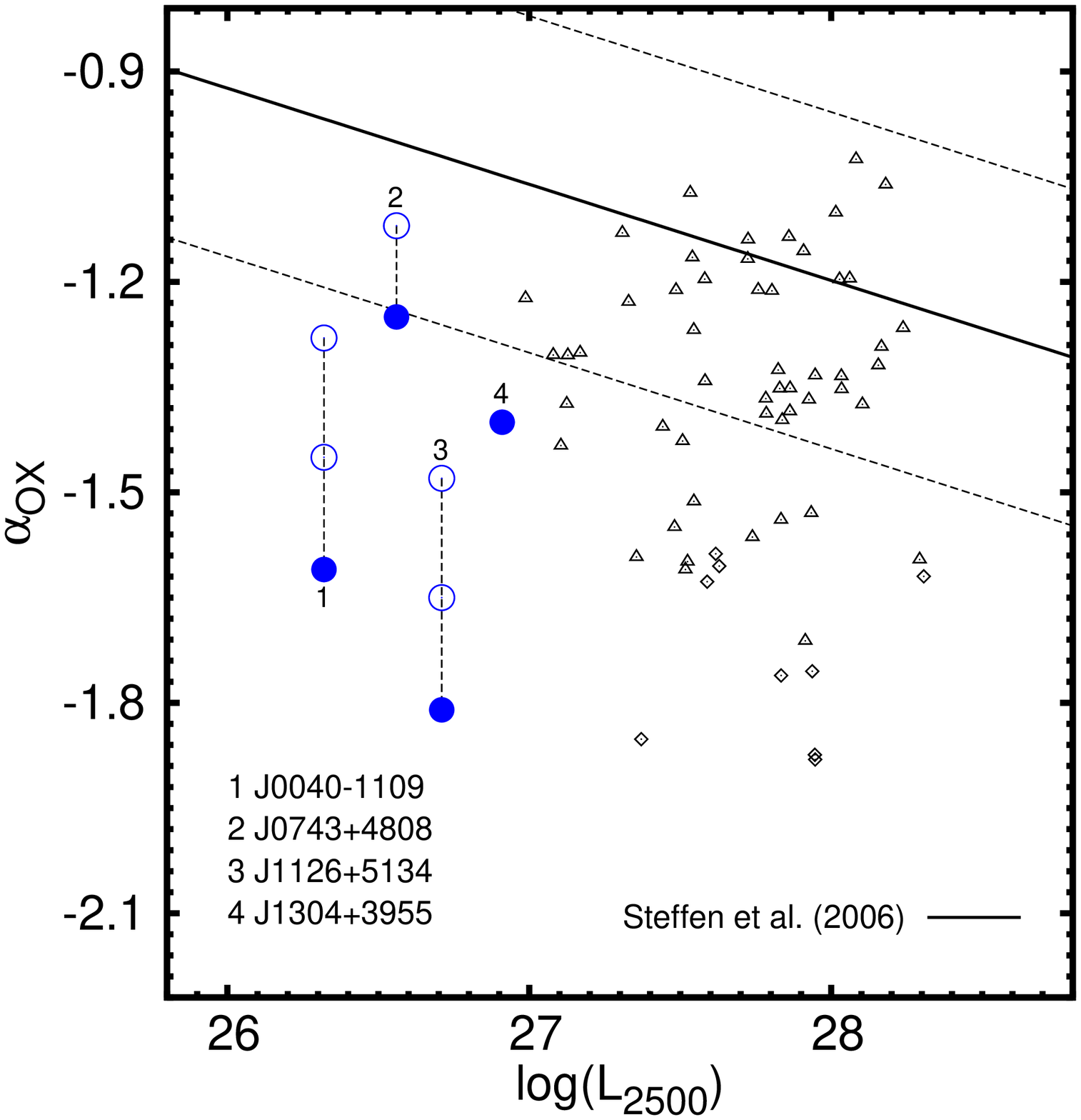}
\caption{
X-ray luminosity vs.\ the [OIII] line luminosity  (left-hand panel) and
the optical-to-X-ray spectral index \aox\ 
vs.\ the 2500\AA\ luminosity (right-hand panel)
for our sample objects;
filled dots: the X-ray luminosities for 
an unabsorbed power-law spectral model; 
open circles: the X-ray luminosities corrected for assumed intrinsic 
absorption which have the absorber parameters the same as derived for
J0743+4808 (see text). 
For the two X-ray non-detections 
(J0040$-$1109 and J1126+5134), the values represent upper limits;
furthermore, an even more conservative case is also overplotted where 
a full absorption coverage is assumed
(corresponding to the case of the highest intrinsic X-ray luminosity).
The low-mass AGN objects with 
high Eddington ratios observed with Chandra from the \citet{dgh12} sample 
are also plotted for comparisons (triangles: X-ray detections;
diamond: X-ray non-detections).
The solid line in the left panel represents the relation given by 
Panessa et al.\ (2006); the solid line and dotted lines 
in the right panel represent
the relation and the 68\% scatter, respectively, given by Steffen et al.\ (2006).
} 
\label{fig:lxlo3}
\end{figure}

Here the multi-wavelength properties of these under-luminous low-mass AGNs
are investigated and compared to other AGN samples.
The X-ray and [OIII] line luminosity relation is shown in
Figure\,\ref{fig:lxlo3} (left panel); 
also plotted is the 
low-mass AGN sample with high Eddington ratios ($\geq 1$) from \citet{dgh12}. 
For J0040-1109 and J1126+5134, 
the upper limits on the X-ray luminosities are shown, 
which correspond to the three cases from the most 
stringent to the most conservative limits:
(i) assuming a power-law spectrum with $\Gamma=1.7$ without intrinsic absorption (filled dots);
(ii) corrected for  assumed  absorption of  partial covering  as discussed in 
Section\,\ref{sect:er} (open circles);
(iii) the same as (ii) but assuming fully covering absorber (open circles).
As can be seen, our objects are systematically fainter than 
their high-accretion counterparts, 
and extend the previously known \lx--\loiii\ relation down to 
the lowest luminosity regime ever probed for AGNs.
The trend is roughly consistent with the extrapolation of the 
previously known relations 
for classical AGNs with high BH masses
\citep[e.g.][]{pan06},  albeit with large scatters.
The scatter may be contributed from a few sources:
for instance, X-ray variability, 
as seen from this work and other observations of low-mass AGNs
\citep{vaug05,dew08,min09,ai11};
or possible absorption of the X-ray emission in some objects. 
For J1126+5134, its X-ray weakness may be at least
partly ascribed to absorption,
as its deviation from the \lx--\loiii\ relation gets smaller 
once the luminosity limit is corrected for absorption 
(with \nh\ on the order of $10^{22}$\,\unh). 
On the contrary, little or no X-ray absorption is inferred in J0040-1109,
suggestive of an extremely low intrinsic X-ray luminosity
that is comparable or even lower than that of NGC\,4359---the least
luminous dwarf AGN known so far.

The optical/UV ($2500$\AA) to X-ray (2\,keV) effective spectral indices
\aox\  are calculated (Table\,\ref{tab:xlum}),  
which is defined as the two-point effective spectral index 
between $2500$\AA\ and 2\,keV\footnote{\aox$\equiv {\rm log}(f_{\rm 2keV}/f_{\rm 2500A})/{\rm log}(\nu_{\rm 2keV}/\nu_{\rm 2500A})= 0.384 {\rm log}(f_{\rm 2keV}/f_{\rm 2500A})$
}.
The $2500\AA$ luminosity is calculated from the $5100$\AA\ luminosity which
is estimated from the broad \ha\ line assuming an  
optical spectral index $\alpha= -0.5$ ($S\propto \nu^{\alpha}$). 
For the X-ray non-detections
the lower limits are calculated from the nominal X-ray luminosity limits.
Their \aox\ values are plotted versus the $2500$\AA\ luminosities 
in Figure\,\ref{fig:lxlo3} (right panel).
For J0040-1109 and J1126+5134, 
we adopt the \aox\ limits corresponding to the X-ray luminosity limits 
shown in the left panel of the figure, i.e.\ considering the cases of X-ray absorption  (see above).
It shows that our objects are comparable to the 
luminous low-mass AGN sample of \citet{dgh12} in the \aox\ distribution,
which has a large scatter. 
It also shows that our objects, being one order of magnitude fainter in \lomon\
than the \citet{dgh12} sample, do not follow the extrapolation of the
previously found \aox--\lomon\ relation, such as that in \citet{ste06}
(see Yuan et al.\ 1998 for a different view, however).
In particular, the two non-detections appear to be `X-ray weak', 
with respect to their optical/UV emission.
This is possibly true for J0040-1109, 
which requires little or no X-ray absorption,
whereas  X-ray absorption might play a role 
in J1126+5134, however,
as argued above based on their \loiii\ luminosities.

The deviation from the empirical \aox-\lomon\ relation at low \lomon\ 
is not surprising, as theoretically \aox\ should directly depend on both
\mbh\ and accretion rates, rather than \lomon.
It should be noted that 
this deviation would be aggravated if the optical light 
suffers from dust extinction, though we do not consider it to be significant.  
The steep \aox\ values are not unexpected at low Eddington ratios
in accretion disk emission models \citep[e.g.][]{done12}, 
despite the somewhat ad hoc modeling of the X-ray emission.
This is simply a consequence of that 
the peak of the multi-temperature blackbody spectrum 
of the accretion disk is shifted toward low energies with decreasing accretion rates.
(see Figure\,6 of Dong R.\ et al. 2012 for discussion on these dependences). 
Furthermore, variability may also play a role, 
as the X-ray and optical observations were not simultaneous.
In addition to the expected large amplitude X-ray variability
for their small BH masses, recent observations show that, statistically, 
long-term optical variability is increasing with
a decrease of the Eddington ratios in AGNs \citep{ai10}.

\subsection{A Large Population of Low-mass BH in the Local Universe?}

The Eddington ratios of  our  objects are close to the critical accretion rate
in Galactic BH X-ray binaries, typically around 2\%--3\% of the Eddington accretion rate, 
below which a standard thin disc is 
truncated with its inner region replaced by an
advection dominated accretion flow \citep{ny94}. 
Theoretical studies show that this process is also applicable in AGNs
\citep{lm01}.
AGNs with lower Eddington ratios than the critical rate
might undergo such an inner disk truncation.
This would result in a reduced UV continuum and hence weaker emission lines, 
making their line spectra even less prominent against the contamination of 
host galaxy starlight and, hence, difficult to detect.
Moreover, absorption/obscuration 
would make the objects dimmer in the X-ray and/or optical bands.
It has been suggested that the fraction of obscured AGNs 
increases significantly
with decreasing X-ray luminosity \citep[e.g.][]{bur11}.
If this trend holds for low-mass AGNs, 
obscuration must be very common in such objects, 
especially in objects with low \redd.
In fact, as discussed above, 
obscuration of the X-ray emission may be inferred in at least one
 of our objects. 
A combination of these effects would further aggravate 
the elusion of low-mass AGNs from detection in optical (and also X-ray) surveys.

The {\em intrinsic} Eddington ratio distribution function (ERDF) 
for local active BHs (\redd$>10^{-2}$) in the mass range of 
\mbh$=10^{6-9}$\,\msun\  
was derived by \citet{sw10}, 
based on a well-defined sample and taking selection biases into account.
The ERDF  can be described by a Schechter function and 
exhibits a monotonic rise toward the lower end down to \redd$\sim10^{-2}$.
Assuming that the same ERDF holds for low-mass AGNs,
we try to estimate the "intrinsic" number of low-mass AGNs at the lower \redd\
end from the detected objects at the higher \redd\ end;
the latter are comparatively much brighter
and can thus be considered to be more or less complete 
(volume limited with $z<0.35$).
For the high-\redd\ end we choose $\log$(\redd)$>-0.2$
(\redd$>0.63$) as a reasonable value, 
and there are 40 such objects in the D12 sample. 
We find that, by integrating the ERDF, 
objects with $\log$(\redd)$>-0.2$ make up only 
0.06\% 
of all the objects with $\log$(\redd)$>-2$.
This predicts  a total  of at least $6.6\times10^4$ low-mass AGNs with 
$\log$(\redd)$>-2$,  a factor of about 200 more than 
what have actually been detected from  SDSS.
This number would be even larger when extending \redd\ to below $10^{-2}$.
It should be noted that only type\,1 AGNs are considered in the above estimation.
The number ratio of type\,2 to type\,1 AGNs  is not known in this \mbh\ regime. 
For more massive Seyfert galaxies, the ratio shows an increase toward low X-ray luminosities 
and reaches $\sim4$ or even higher at the lowest luminosity \citep[e.g.][]{bur11}.
If this ratio is not  largely different from that for massive Seyfert galaxies, 
the expected number of low-mass AGNs would be even larger by a factor of four or more.
We thus conclude that, 
unless their intrinsic ERDF exhibits a significant drop below 
\redd$\sim0.6$,  a large population of low-mass BHs or 
IMBHs with \mbh$<10^{6}$\,\msun\  probably exists in the local universe.

\section{Summary and implications}

We performed X-ray observations with Chandra of four low-mass AGN candidates
(\mbh$\la10^6$\,\msun)  selected from the SDSS, 
aiming at validating them as AGNs  and having
 low Eddington ratios that were inferred based on broad \ha.
 In addition, for two objects (J0743+4808 and J1126+5134) 
 having the lowest significance of broad \ha,
 optical spectroscopic observations with MMT were performed at better 
 spatial resolution, which confirm them as Seyfert\,1 galaxies 
 and low \mbh.
Of the four dwarf galaxies, X-ray emission from the nuclei of 
two (J0743+4808 and J1304+3955) is detected,
which varies on timescales as short as $\sim 10^3$\,s,
indicating an AGN origin in both.
For the remaining two, the detection is not significant, or only marginal at best.
These results imply that 
 the vast majority, if not all, of the objects in the D12 sample 
are  low-mass Seyfert\,1 galaxies, 
which were selected purely based on their broad \ha\ lines.
Serendipitously, an off-nucleus ULX in the dwarf galaxy J0743+4808 is detected, 
with a 2--10\,keV luminosity of $(6-9)\times10^{39}$\,\ulum\ assuming
typical spectral indices for ULXs, 
which may imply a BH with a mass of at least 
50--70\,\msun\ if it is accreting at the Eddington rate
and radiating isotropically.

The X-ray luminosities of the sample objects are found to range from 
below $10^{40}$\,\ulum\  to a few times $10^{41}$\,\ulum\  in 2--10\,keV.
This luminosity regime is among the lowest ever probed for an AGN, 
and is actually comparable to bright ULXs.
Extending to such extremely low luminosities,
our data are still broadly consistent with the known \lx--\loiii\ relation,
but seem to deviate from the previous \aox--\lomon\ relation.
The two detected objects show diverse X-ray spectral shapes, 
with a power-law photon index $\Gamma$=$1.70\pm0.18$ 
in  J1304+3955 and an apparently unusually flat $\Gamma$=$1.02\pm0.13$ in J0743+4808
(68\% confidence level).
The Eddington ratios range from $\sim 10^{-3}$  to $\sim 10^{-2}$, 
for a range of reasonable X-ray-to-bolometric luminosity corrections
and spectral models.
Thus their low Eddington ratios are confirmed.
This implies that the detection of the
low-\redd, low-mass AGNs in the D12 sample should mostly be reliable.

The low-\redd\ objects found in the D12 sample 
 are already at the detection limit of the SDSS data 
for a low-mass AGN by means of the broad emission line method 
as employed therein.
Low-mass AGNs in dwarf galaxies with even lower \redd, 
or at larger distances,  or residing in brighter nuclear star clusters
would have most likely been missed in the SDSS survey, as generally expected.
Therefore, there likely exist a possibly large population of black holes
with \mbh$=10^{5-6}$\,\msun\ or even less in the local universe.
Future sky surveys at multi-wavelengths, such as {\it eROSITA} in X-ray, 
will reveal more objects of this population.

\acknowledgments
We thank the referee for useful  comments that helped improve this paper. 
This work is supported by the NSFC grant (NSF11033007) and
the National Basic Research Program of (973 Program) 2009CB824800.
W.Y. thanks Luis Ho for comments,  R.\ Dong for providing the data from their paper,
and A.\ Schulze for providing the numerical results of their ERDF.
Funding for the SDSS and SDSS-II has been provided by the Alfred P.
Sloan Foundation, the Participating Institutions, the National
Science Foundation, the U.S. Department of Energy, the National
Aeronautics and Space Administration, the Japanese Monbukagakusho,
the Max Planck Society, and the Higher Education Funding Council for
England.  The SDSS Web site is http://www.sdss.org/. The SDSS is
managed by the Astrophysical Research Consortium for the
Participating Institutions. The Participating Institutions are the
American Museum of Natural History, Astrophysical Institute Potsdam,
University of Basel, University of Cambridge, Case Western Reserve
University, University of Chicago, Drexel University, Fermilab, the
Institute for Advanced Study, the Japan Participation Group, Johns
Hopkins University, the Joint Institute for Nuclear Astrophysics,
the Kavli Institute for Particle Astrophysics and Cosmology, the
Korean Scientist Group, the Chinese Academy of Sciences (LAMOST),
Los Alamos National Laboratory, the Max-Planck-Institute for
Astronomy (MPIA), the Max-Planck-Institute for Astrophysics (MPA),
New Mexico State University, Ohio State University, University of
Pittsburgh, University of Portsmouth, Princeton University, the
United States Naval Observatory, and the University of Washington.
This research has made use of the NASA/IPAC Extragalactic Database (NED) which is operated by the Jet Propulsion Laboratory, California Institute of Technology, under contract with the National Aeronautics and Space Administration.


\begin{deluxetable}{lcccccc}
\tablenum{1} \tablewidth{0pt} \topmargin 0.0cm \evensidemargin = 0mm
\oddsidemargin = 0mm
\tabletypesize{\scriptsize}
\tablecaption{Sample and basic source parameters}
\tablehead{ \colhead{SDSS name} &
            \colhead{redshift} &
            \colhead{FWHM(\ha)} &
            \colhead{L(\ha)} &
            \colhead{\mbh} &
            \colhead{\redd} & 
            \colhead{$M_{\rm B}$}\\
            \colhead{} &
            \colhead{} &
            \colhead{\kmps} &
            \colhead{$10^{39}$\ulum} &
            \colhead{$10^6$\,\msun} &
            \colhead{$10^{-2}$} &
            \colhead{mag} \\            
            \colhead{(1)} &
            \colhead{(2)} &
            \colhead{(3)}&
            \colhead{(4)} &
            \colhead{(5)} &
            \colhead{(6)} &
            \colhead{(7)} 
                        }
\startdata
J004042.10$-$110957.6 & 0.027& 2240 & 3.4 & 1.22& 1.0 & $-17.4$\\
J074345.47+480813.5   & 0.018& 1287(1450)  & 6.5 & 0.51(0.66) & 4.3(3.3) & $-18.0$\\
J112637.74+513423.0   & 0.026& 1481(1875) & 9.7 & 0.83(1.35)& 3.7(2.3) & $-18.9$\\
J130456.95+395529.7   & 0.027& 1413 & 16.5& 0.95& 5.2 & $-19.7$ \\
\enddata
\tablecomments{
Values given in brackets are measurements from the MMT observations.
Col. (1): SDSS names;
Col. (2): redshift;
Col. (3): linewidth of the broad \ha\ component;
Col. (4): luminosity of the broad \ha\ line;
Col. (5): black hole mass;
Col. (6): Eddington ratio;
Col. (7): $B$-band absolute magnitude of host galaxy
}
\label{tab:sample}
\end{deluxetable}

\begin{deluxetable}{lcccccc}
\tablenum{2} \tablewidth{0pt} \topmargin 0.0cm \evensidemargin = 0mm
\oddsidemargin = 0mm
\tabletypesize{\scriptsize}
\tablecaption{X-ray observation logs and source detection}
\tablehead{ \colhead{short name} &
            \colhead{obs-ID} &
            \colhead{date} &
            \colhead{\gnh} &
            \colhead{$T_{\rm expo}$}  &
            \colhead{cts} &
            \colhead{CR} \\
            \colhead{} &
            \colhead{} &
            \colhead{} &
            \colhead{$10^{20}$\,\unh} &
            \colhead{sec}  &
            \colhead{} &
            \colhead{$10^{-3}$\,\ucr} \\
            \colhead{(1)} &
            \colhead{(2)} &
            \colhead{(3)}&
            \colhead{(4)} &
            \colhead{(5)} &
            \colhead{(6)} &
            \colhead{(7)} 
            }
\startdata
J0040$-$1109   & 09235 & 2007-12-27 & 2.37 & 4686 & $<6.4$& $<1.4$ \\
J0743+4808     & 09233 & 2007-12-14 & 5.40 & 4675 & 194   & 41.5  \\
J0743+4808 X-1 & 09233 & 2007-12-14 & 5.40 & 4675 & 12    &  2.3   \\
J1126+5134     & 09234 & 2008-03-26 & 1.18 & 4667 & $<5.2$& $<1.1$\\
J1304+3955     & 09236 & 2008-04-05 & 1.46 & 4699 & 94    &  20.0 \\
\enddata
\tablecomments{
Col. (1): object names; J0743+4808 X-1 denotes the off-nucleus X-ray source
in J0743+4808; 
Col. (2): Chandra observation ID;
Col. (3): observational date;
Col. (4): Galactic HI column density;
Col. (5): exposure time;
Col. (6): source counts in 0.3--8\,keV (upper limits are at the 90\% confidence level);
Col. (7): count rate in 0.3--8\,keV.
}
\label{tab:xobs} 
\end{deluxetable}

\begin{deluxetable}{lccccccc}
\tablenum{3} \tablewidth{0pt} \topmargin 0.0cm \evensidemargin = 0mm
\oddsidemargin = 0mm
\tabletypesize{\scriptsize}
\tablecaption{
Results of X-ray spectral fitting for J0743+4808
}
\tablehead{ \colhead{wabs * model} &
            \colhead{$\Gamma$} &
            \colhead{$f_{\rm 2-10\,keV}$} &
            \colhead{\nh} &            
            \colhead{$\log\xi$} &
            \colhead{$f_{\rm c}$ } &
            \colhead{$\chi^2$/d.o.f.} &
            \colhead{$P_{\rm null}$}\\
            \colhead{} &
            \colhead{} &
            \colhead{$10^{-13}$\ergs} &
            \colhead{$10^{22}$\,\unh} &
            \colhead{} &
            \colhead{} &
            \colhead{} &
            \colhead{} \\
            \colhead{(1)} &
            \colhead{(2)} &
            \colhead{(3)} &
            \colhead{(4)} &
            \colhead{(5)} &
            \colhead{(6)} &
            \colhead{(7)} &
            \colhead{(8)}
            }                                             
\startdata
power-law          & $1.02\pm 0.13$  & $5.2^{+0.9}_{-0.8}$ &   --  & -- & -- & 8.4/6  & 0.21\\
power-law          & 1.7 (fixed)     &  --   &   --  & -- & -- &  29/7  & $1.4\times 10^{-4}$\\
PL*zxipcf   & $1.31^{+0.21}_{-0.24}$& $5.8\pm1.1$ & $4.2^{+3.7}_{-2.8}$ & $2.4^{+0.6}_{-0.4}$ & 1 (fixed) & 4.4/4 & 0.35\\ 
PL*zxipcf   & 1.7 (fixed)  & $4.6\pm0.7$  & $1.9^{+0.8}_{-0.7}$ & $1.5^{+0.3}_{-0.3}$ & 1 (fixed) & 6.8/5 & 0.23 \\
PL*zxipcf   & 1.7 (fixed)  & $6.1\pm1.4$           & $5.4^{+2.7}_{-2.3}$ & $1.3^{+0.4}_{-1.5}$ & $0.78^{+0.10}_{-0.09}$ & 5.3/4 & 0.26 \\
PL*zxipcf   & 1.7 (fixed)  & $6.5^{+1.6}_{-1.8}$ & $1.8^{+0.9}_{-1.0}$ & -3 (fixed)          & $0.73^{+0.06}_{-0.09}$ & 6.5/5 & 0.26 \\
\enddata
\tablecomments{
Col.(1) XSPEC models; wabs: absorption with column density fixed at the Galactic value;
PL: power-law; zxipcf: partial covering photoionized absorption based on XSTAR;  
(2) photon index of power-law spectrum ($f(E)=KE^{-\Gamma}$);
(3) flux density (2--10\,keV) in the observer's frame corrected for Galactic and intrinsic (if any) absorption;
(4) intrinsic absorption column density;
(5) ionization parameter;
(6) covering fraction; 
(7) $\chi^2$ and degree of freedom of the fits;
(8) corresponding null probability for the $\chi^2$.
}
\label{tab:xfit}
\end{deluxetable}

\begin{deluxetable}{lccc}
\tablenum{4} \tablewidth{0pt} \topmargin 0.0cm \evensidemargin = 0mm
\oddsidemargin = 0mm
\tabletypesize{\scriptsize}
\tablecaption{X-ray luminosities and X-ray to optical indices}
\tablehead{ \colhead{name} &
            \colhead{$L_{\rm 2-10keV}$} &
            \colhead{$L_{\rm 2-10keV}/L_{\rm Edd}$} &
            \colhead{\aox} \\
            \colhead{} &
            \colhead{$10^{40}$\,\ulum} &
            \colhead{} &
            \colhead{}  \\
            \colhead{(1)} &
            \colhead{(2)} &
            \colhead{(3)}&
            \colhead{(4)} 
            }
\startdata
J0040$-$1109   & $<1.3(2.8)$ & $<0.8(1.7)\times10^{-4}$ & $<$-1.61(-1.48) \\
J0743+4808     & $37.6 (44.7)$   & $ 4.4(5.2)\times 10^{-3}$    & -1.25(-1.12)  \\
J1126+5134     & $<1.0(2.1)$ & $<0.6(1.2)\times10^{-4}$ & $<$-1.81(-1.68)\\
J1304+3955     & $18.4$           & $1.5\times10^{-3}$ & -1.40 
\enddata
\tablecomments{
Col.(1): object names;
(2): 2--10\,keV luminosity (corrected for absorption, if any);
(3): ratio of 2--10\,keV luminosity to the Eddington luminosity;
(4): optical to X-ray effective spectral index.
For J0743+4808,  the values preceding brackets are for the case of 
a power-law model, and 
those in brackets correspond to the case corrected for the 
intrinsic absorption (model No.\,5 in Table\,\ref{tab:xfit}).
In the cases of J0040$-$1109 and J1126+5134 (X-ray non-detections),
the values preceding brackets are for the case of an assumed power-law model with 
the photon index $\Gamma=1.7$, 
and those in brackets are for  $\Gamma=1.0$.
}
\label{tab:xlum} 
\end{deluxetable}


\begin{thebibliography}{}

\bibitem[Ai et al.(2011)]{ai11} Ai, Y.L., Yuan, W., Zhou, H.Y., Wang, T.G., Zhang, S.H. 2011, \apj, 727, 31

\bibitem[Ai et al.(2010)]{ai10} Ai, Y.L., Yuan, W., Zhou, H.Y., Wang, T.G., Dong, X.-B., et al.\ 2010, \apj, 716, L31

\bibitem[Barth et al.(2004)]{bar04} Barth, A.~J., Ho, L.~C., Rutledge, R.~E., \& Sargent, W.~L.~W.\ 2004, \apj, 607, 90

\bibitem[Bevington \& Robinson(1992)]{bev92} Bevington, P.R. \& Robinson, D.K. 1992, Data reduction and error analysis for the physical sciences, 2nd ed. (WCB/McGraw-Hill) 

\bibitem[Burlon et al.(2011)]{bur11} Burlon, D., Ajello, M., Greiner, J., Comastri, A., Merloni, A., Gehrels, N. 2011, \apj, 728, 58

\bibitem[Czerny(2006)]{cze06}Czerny, B. 2006, in ASP Conference Series, 360 AGN Variability from X-Rays to Radio Waves, ed. C. M. Gaskell et al. (San Francisco, CA:ASP), 265

\bibitem[Desroches et al.(2009)]{des09} Desroches, L.-B., Greene, J. E., \& Ho, L. C. 2009, \apj, 698, 1515

\bibitem[Desroches \& Ho(2009)]{dh09} Desroches, L.-B. \& Ho, L.C. 2009, \apj, 690, 267

\bibitem[Dewangan et al.(2008)]{dew08} Dewangan, G.C., Mathur, S., Griffiths, R.E., Rao, A.R. 2008, \apj, 689, 762

\bibitem[Done et al.(2012)]{done12} Done, C., Davis, S.W., Jin, C., Blaes, O., \& Ward, M. 2012, \mnras, 420, 1848

\bibitem[Dong R. et al.(2012)]{dgh12} Dong, R., Greene, J.E. \& Ho, L.C. 2012, \apj, 761, 73


\bibitem[Dong X.-B. et al.(2012)]{dong12} Dong, X.-B., Ho, L.C., Yuan, W., Wang, T.-G., Fan, X., Zhou, H., Jiang, N.  2012, \apj, 755, 167 

\bibitem[Dong X.-B. et al.(2007)]{dong07} Dong, X.-B., Wang, T.-G., Yuan W., et al.\ 2007, \apj, 657, 700

\bibitem[Filippenko \& Ho(2003)]{fh03} Filippenko, A. V., \& Ho, L. C. 2003, \apj, 588, L13

\bibitem[Filippenko et al.(1993)]{fhs93} Filippenko, A.~V., Ho, L.~C. \& Sargent, W.~L.~W.\ 1993, \apjl, 410, L75

\bibitem[Fitzpatrick(1999)]{fit99} Fitzpatrick, E.~L.\ 1999, \pasp, 111, 63

\bibitem[Greene \& Ho(2004)]{gh04} Greene, J. E., \&   Ho, L. C. 2004, \apj, 610, 722


\bibitem[Greene \& Ho(2005)]{gh05b} Greene, J.~E. \& Ho, L.~C.\ 2005, \apj, 630, 122

\bibitem[Greene \& Ho(2007a)]{gh07a} Greene, J. E. \&   Ho, L. C. 2007a, \apj, 656, 84

\bibitem[Greene \& Ho(2007b)]{gh07b} Greene, J. E. \&   Ho, L. C. 2007b, \apj, 670, 92

\bibitem[Haehnelt et al.(1998)]{hae98} Haehnelt, M. G., Natarajan, P. \& Rees, M. J.\ 1998, \mnras, 300, 817

\bibitem[Ho(2009)]{ho09} Ho, L.C. 2009, \apj, 699, 626

\bibitem[Iwasawa et al.(2000)]{iwa00} Iwasawa, K., Fabian, A.C., Almanini, O., et al. 2000, \mnras, 318, 879

\bibitem[Iwasawa et al.(2010)]{iwa10} Iwasawa, K., Tanaka, Y., Gallo, L.C. 2010, \aap, 514, 58

\bibitem[Lira et al.(1999)]{lira99} Lira, P., Lawrence, A., OÕBrien, P., et al.\ 1999, \mnras, 305, 109

\bibitem[Liu \& Meyer-Hofmeister(2001)]{lm01} Liu, B.F. \& Meyer-Hofmeister, E.\ 2001, \aap, 372, 386

\bibitem[Lu et al.(2006)]{lu06} Lu, H., Zhou, H., Wang, J., Wang, T., Dong, X., Zhuang, Z., \& Li, C.\ 2006, \aj, 131, 790

\bibitem[Mateos et al.(2010)]{mat10} Mateos, S., Carrera, F.J., Page, M.J., Watson, M.G., Corral, A., et al.\  2010, \aap, 510, 35

\bibitem[Miniutti et al.(2009)]{min09}Miniutti, G., Ponti, G., Greene, J. E., Ho, L. C., Fabian, A. C., \& Iwasawa, K., 2009, \mnras, 394, 443

\bibitem[McLure \& Dunlop(2004)]{md04} McLure, R.~J. \& Dunlop, J.~S.\ 2004, \mnras, 352, 1390  

\bibitem[Moran et al.(2005)]{mor05} Moran, E.C., et al.\ 2005, \aj, 129, 2108

\bibitem[Mushotzky et al.(1993)]{mus93} Mushotzky, R.F., Done, C., Pounds, K.A.\ 1993, \araa, 31, 717

\bibitem[Narayan \& Yi(1994)]{ny94} Narayan, R. \& Yi, I.\ 1994, \apj, 428, L13

\bibitem[Nardini \& Risaliti(2011)]{nr11} Nardini, E. \& Risaliti, G.\ 2011 \mnras, 417, 2571

\bibitem[Panessa et al.(2006)]{pan06} Panessa, F., Bassani, L., Cappi, M., et al.\ 2006, \aap, 455, 173

\bibitem[Peterson et al.(2005)]{pet05} Peterson, B.M., et al.\ 2005, \apj, 632, 799

\bibitem[Risaliti et al.(2009)]{ris09} Risaliti, G., Young, M., \& Elvis, M. 2009, \apj, 700, 6

\bibitem[Satyapal et al.(2008)]{sat08} Satyapal, S., Vega, D., Dudik, R. P., Abel, N. P., \& Heckman, T. 2008, \apj, 677, 926

\bibitem[Schlegel et al.(1998)]{sfd98} Schlegel, D.~J., Finkbeiner, D.~P. \& Davis, M.\ 1998, \apj, 500, 525

\bibitem[Schulze \& Wisotzki (2010)]{sw10} Schulze, A. \& Wisotzki, L.\ 2010 \aap, 516, 87 

\bibitem[Shih et al.(2003)]{shi03} Shih, D.C., Iwasawa, K. \& Fabian, A.C. 2003, \mnras, 341, 973

\bibitem[Singh et al.(2011)]{sing11} Singh, V., Shastri1, P. \& Risaliti, G.\ et al. 2011 \aap, 533, 128

\bibitem[Stefen et al.(2006)]{ste06} Steffen, A.T., Strateva, I., Brandt, W.N., et al. 2006, \aj, 131, 2826

\bibitem[Thornton et al.(2009)]{tho09}  Thornton, C.E., Barth, A.J., Ho, L.C., Greene, J.E. 2009, \apj, 705, 1196

\bibitem[Vasudevan \& Fabian (2007)]{vf07} Vasudevan, R.V. \& Fabian, A.C.\ 2007, \mnras, 381, 1235

\bibitem[Vasudevan \& Fabian (2009)]{vf09} Vasudevan, R.V. \& Fabian, A.C. 2009, \mnras, 392, 1124

\bibitem[Vasudevan et al.(2009)]{vas09} Vasudevan, R.V., Mushotzky, R.F., Winter, L.M. \& Fabian, A.C. 2009, \mnras, 399, 1553

\bibitem[Vaughan et al.(2005)]{vaug05} Vaughan, S., Iwasawa, K, Fabian, A.C., Hayashida, K.\ 2005, \mnras\ 356, 524

\bibitem[Vignali et al.(2001)]{vig01} Vignali, C., Brandt, W. N., Fan, X., Gunn, J. E., Kaspi, S., et al. 2001, \aj, 122, 2143

\bibitem[Volonteri et al.(2008)]{vol08} Volonteri, M., Lodato, G., \& Natarajan, P.\ 2008, \mnras, 383, 1079

\bibitem[Yuan et al.(1998)]{yuan98} Yuan, W., Siebert, J., Brinkmann, W.\ 1998, \aap, 334, 498

\bibitem[Zhou et al.(2006)]{zhou06} Zhou, H., Wang, T.-G., Yuan,
W., Lu, H., Dong, X.-B., Wang, J., \& Lu, Y.\ 2006, \apjs, 166, 128

\end{thebibliography}
\end{document}